\documentclass[preprint,12pt,3p]{elsarticle}

\usepackage{amssymb}
\usepackage{amsmath}
\usepackage{multirow}
\usepackage{microtype}

\usepackage{lineno}

\usepackage{siunitx}
\usepackage{caption}
\usepackage{subcaption}
\usepackage{url}
\usepackage{hyperref}

\usepackage{ marvosym }

\usepackage{bits}

\newcommand{\vlvnt}{VLV\ensuremath{\nu}T}
\newcommand{\nue}{\ensuremath{\nu_e}}
\newcommand{\numu}{\ensuremath{\nu_\mu}}
\newcommand{\nutau}{\ensuremath{\nu_\tau}}
\newcommand{\nuebar}{\ensuremath{\bar{\nu}_e}}
\newcommand{\numubar}{\ensuremath{\bar{\nu}_\mu}}
\newcommand{\nutaubar}{\ensuremath{\bar{\nu}_\tau}}

\newcommand{\numucc}{\ensuremath{\numu\rm{\ CC}}}

\newcommand{\numubarcc}{\ensuremath{\numubar\rm{\ CC}}}

\newcommand{\Etrue}{\ensuremath{E_{\rm true}}}
\newcommand{\Ereco}{\ensuremath{E_{\rm reco}}}
\newcommand{\CZtrue}{\ensuremath{\cos{\vartheta}_{\rm true}}}
\newcommand{\CZreco}{\ensuremath{\cos{\vartheta}_{\rm reco}}}

\journal{Journal}

\begin{document}

\begin{frontmatter}

\title{Computational Techniques for the Analysis of Small Signals in High-Statistics Neutrino Oscillation Experiments}

\author[christchurch]{M. G. Aartsen}
\author[zeuthen]{M. Ackermann}
\author[christchurch]{J. Adams}
\author[brusselslibre]{J. A. Aguilar}
\author[copenhagen]{M. Ahlers}
\author[stockholmokc]{M. Ahrens}
\author[geneva]{C. Alispach}
\author[marquette]{K. Andeen}
\author[pennphys]{T. Anderson}
\author[brusselslibre]{I. Ansseau}
% ...
\address[brusselslibre]{Universit{\'e} Libre de Bruxelles, Science Faculty CP230, B-1050 Brussels, Belgium}
% ...
\address[christchurch]{Dept. of Physics and Astronomy, University of Canterbury, Private Bag 4800, Christchurch, New Zealand}
% ...
\address[copenhagen]{Niels Bohr Institute, University of Copenhagen, DK-2100 Copenhagen, Denmark}
% ...
\address[geneva]{D{\'e}partement de physique nucl{\'e}aire et corpusculaire, Universit{\'e} de Gen{\`e}ve, CH-1211 Gen{\`e}ve, Switzerland}
% ...
\address[marquette]{Department of Physics, Marquette University, Milwaukee, WI, 53201, USA}
% ...
\address[stockholmokc]{Oskar Klein Centre and Dept. of Physics, Stockholm University, SE-10691 Stockholm, Sweden}
% ...
\address[pennphys]{Dept. of Physics, Pennsylvania State University, University Park, PA 16802, USA}
% ...
\address[zeuthen]{DESY, D-15738 Zeuthen, Germany}
\fntext[padova]{also at Universit{\`a} di Padova, I-35131 Padova, Italy}
\fntext[mephi]{also at National Research Nuclear University, Moscow Engineering Physics Institute (MEPhI), Moscow 115409, Russia}
\fntext[tokyofn]{Earthquake Research Institute, University of Tokyo, Bunkyo, Tokyo 113-0032, Japan}

\cortext[cor]{analysis@icecube.wisc.edu}

\begin{abstract}

The current and upcoming generation of Very Large Volume Neutrino Telescopes---collecting unprecedented quantities of neutrino events---can be used to explore subtle effects in oscillation physics, such as (but not restricted to) the neutrino mass ordering.
The sensitivity of an experiment to these effects can be estimated from Monte Carlo simulations.
With the high number of events that will be collected, there is a trade-off between the computational expense of running such simulations and the inherent statistical uncertainty in the determined values.
In such a scenario, it becomes impractical to produce and use adequately-sized sets of simulated events with traditional methods, such as Monte Carlo weighting. In this work we present a staged approach to the generation of binned event distributions in order to overcome these challenges. By combining multiple integration and smoothing techniques which address limited statistics from simulation it arrives at reliable analysis results using modest computational resources.

\end{abstract}


\end{frontmatter}




\section{Introduction}
\label{sec:intro}

By virtue of their multi-megaton effective mass paired with the magnitude of the atmospheric neutrino flux, the next generation of Very Large Volume Neutrino Telescopes ({\vlvnt}s) dedicated to neutrino oscillation physics, such as the IceCube Upgrade, PINGU, and ORCA~\cite{upgrade_icrc:2019,TheIceCube-Gen2:2016cap,PINGU-LOI,Adrian-Martinez:2016fdl}, will record tens of thousands of GeV-scale neutrino interactions.
These large-scale water or ice Cherenkov detectors do not have the ability to unambiguously distinguish between neutrino flavors and interaction types on an event-by-event basis.
Even so, their high statistics data samples can be used to explore effects that are small compared to the background, such as the tau neutrino appearance rate,
the ordering of the neutrino mass eigenstates (NMO), or potential neutrino physics beyond the Standard Model.

All such physics analyses are carried out by comparing the observed event distributions with predictions (hereafter referred to as {\it templates}) obtained from Monte Carlo (MC) simulations.
The physical phenomena
listed above will appear as statistical (in)compatibilities of templates with differences in event counts as small as a few percent.
An inherent problem when trying to quantify these deviations in high-statistics data sets is that the templates must be described with an accuracy better than the magnitude of the effect being investigated.
A limiting factor to the accuracy is the amount of MC simulation available, which is in turn constrained by the availability of computing resources.
This particularly applies during the design optimization phase of a planned experiment, which entails performance assessments of multiple detector variants.

With an adequate machinery at hand to produce templates, extracting the relevant physical and systematic parameters typically proceeds via maximizing the likelihood of obtaining the observed data under a given hypothesis.
A common feature to all statistical methods is that the templates need to be generated for a multitude of parameter combinations, often thousands or even millions.
This process needs to be accurate, but also fast, which typically prohibits the reproduction of the full MC sample for each template.

In this article, we present an approach that allows for the fast creation of accurate templates even from MC sets that are several orders of magnitude smaller than those necessary when using simpler methods.
An alternative approach that does not remove template inaccuracies but rather mitigates their impact on statistical inference is the inclusion of the inherent MC uncertainty in the fit statistic; recent overviews can be found in~\cite{Glusenkamp:2019uir, Arguelles:2019izp}.

Our approach was used to calculate the expected sensitivities for atmospheric neutrino oscillation analyses with the proposed PINGU experiment~\cite{TheIceCube-Gen2:2016cap,PINGU-LOI}, and a similar approach was taken in low-energy sensitivity studies for the KM3NeT design~\cite{Adrian-Martinez:2016fdl}.
Throughout this article, we will use the NMO analysis for a generic {\vlvnt} as an example to illustrate our methods, though it is applicable in a wider range of atmospheric neutrino oscillation analyses, and, in parts and with limitations, to other experiments.
Section~\ref{sec:prob} details the computational challenge at hand, followed by a brief introduction of the example NMO analysis in Section~\ref{sec:nmo}. Our approach to overcome this challenge is presented in Section~\ref{sec:stage overview} and Section~\ref{sec:implementation}, followed by a discussion of the validity of the approach in Section~\ref{sec:valid}. The performance is compared to other typical analysis methods in Section~\ref{sec:res}, while the computational burden is discussed in Section~\ref{sec:bench}. Section~\ref{sec:summary} concludes with a brief summary of the article.
Finally, in \ref{sec:toy} we provide details about the {\vlvnt} toy model that we use to benchmark the performance of all considered analysis approaches.

\section{Computational Challenge}
\label{sec:prob}

The statistical comparison between experimental data and parametric or MC-based predictions allows inference of the values of physics parameters under study. It typically proceeds via a likelihood analysis.
We first discuss its most general concepts and variants, then detail the computational requirements on MC generation, and finally outline two standard methods of mitigating these computational burdens.

\subsection{Likelihood Analysis}
Different types of likelihood analyses in particle physics share common features\footnote{See, for example, \cite{Barlow:2003cx} for a more complete overview.}.
An experiment records data which are used to reconstruct any observables expected to carry the imprint of the physical phenomenon under study.
A selection (triggering, filtering, etc.) is applied in order to enhance the sought signal.
Before performing statistical inference, we need a theoretical model of the observable distributions to compare to the data.
Often this includes complicated processes like particle interactions and detector response that require the use of MC methods.
Hence, not only the data, but also the model is subject to statistical fluctuations.
However, once an appropriate amount of MC events is available, the data $x_i$ can be compared to templates---theoretical distributions---for different physics parameter values $\boldsymbol{\theta}$ via a likelihood function, $L(x_1, x_2, ..., x_n|\boldsymbol{\theta}) = \Pi_i P(x_i|\boldsymbol{\theta})$, where $P(x_i|\boldsymbol{\theta})$ is the probability to observe the data $x_i$ assuming that $\boldsymbol{\theta}$ corresponds to given values of the physics parameters\footnote{If the total number of measurements, $n$, is also a random quantity, the likelihood function can be extended to include the distribution of $n$~\cite{cowan}.}.
The goal is (in the frequentist picture) to find the maximum likelihood estimators (MLEs) $\boldsymbol{\hat{\theta}}$, i.e., the parameter values which maximize $L$.

The methods presented in this paper depend on a likelihood function applied to binned data.  One usually employs either a Poisson likelihood or a $\chi^2$ approximation thereof (see for example \cite{cowan}); our methods are independent of this choice, but we use the latter in the example presented in this article.
Binning the data hides physics signatures smaller than the bin size and thus introduces a loss in sensitivity. This can be mitigated by reducing bin sizes, but smaller bins come at the cost of reduced---and possibly insufficient---MC statistics in each bin.

Apart from the physics parameters of interest, a model often comes with nuisance parameters that are also included in the likelihood function.
This further increases the dimensionality of the MLE search, which relies on numerical routines for multidimensional optimization problems.
For the NMO studies, we use the L-BFGS-B algorithm~\cite{l-bfgs-b:1995} in a $D=8$ dimensional parameter space (see Table~\ref{table:params}).
The number of steps necessary for the optimization to converge depends on the particular analysis and model being used (i.e., the details of the resulting likelihood landscape); in the case of our toy example, an average of $\sim10^3$ templates (one per realization of $\boldsymbol{\theta}$) were needed to converge.

\subsection{Template and MC Generation Requirements}
\label{subsec: MC gen requirements}
The problems associated with generating such a large number of templates are exacerbated when estimating the median sensitivity of an experiment.
The above process needs to be applied to an ensemble of random toy MC pseudo-experiments\footnote{Each pseudo-experiment corresponds to a statistical fluctuation of the expected experimental outcome as predicted by MC events. For certain problems, the generation of pseudo-experiments can be skipped by applying the {\it Asimov} approximation~\cite{Cowan:2010js,PINGU-LOI}.} of size $N_{\rm p}$.
The comparison of test statistic distributions $\mathcal{T}$ (see Section \ref{sec:nmo} for details) can be used to estimate a significance value $n_\sigma$ at which one hypothesis is preferred over the alternative.
If $\mathcal{T}$ is Gaussian distributed\footnote{While not a prediction from the model, a near-Gaussian distribution of the test statistic is observed in most NMO studies~\cite{PINGU-LOI,Adrian-Martinez:2016fdl,Blennow:2014fk}.}, the uncertainty $\Delta n_\sigma$ to which $n_\sigma$ can be determined depends upon the number of pseudo-experiments $N_{\rm p}$ as (see \ref{sec:signif} for details):
\begin{equation}
\label{eq:boeser}
    \Delta n_\sigma = \frac{1}{\sqrt{N_{\rm p}}} \sqrt{\frac{n_\sigma^2}{2} + 2}.
\end{equation}

With an absolute uncertainty $\Delta n_\sigma$ at the \SI{1}{\percent} level, determining the sensitivity of an experiment at a confidence level of \SI{99.7}{\percent} (corresponding to $n_\sigma = 3$) requires $\mathcal{O}(10^4)$ pseudo-experiments.


Finally, the event count expectations, $\boldsymbol{\mu}$, for all bins in the templates must be determined at the same level as the physics effects being investigated, which requires at least $\frac{1}{(\SI{1}{\percent})^2}  = 10^4$ MC events per bin to study sub-percent variations arising in a comparison of the two NMO realizations.
At the same time, the number of bins used in any histograms must be commensurate with the experimental resolution and the feature size of the effect under study.
In the example case, at least $\mathcal{O}(10^{3})$ bins are required to resolve the distinct features of the NMO signature; otherwise the analysis cannot exploit the full potential of the experiment.

Therefore, the brute-force approach to our example case requires a very large number of neutrino events to be simulated: $\mathcal{O}(10^7)$ events for each of $\mathcal{O}(10^3)$ values of $\boldsymbol{\theta}$ probed during the optimization process for each of $\mathcal{O}(10^4)$ pseudo-experiments---a grand total of $\mathcal{O}(10^{14})$ events.
Even if the time to simulate and reconstruct a single event is $1\un{s}$ (a very optimistic estimate for our experiment), full fits to all pseudo-experiments under the two ordering hypotheses would require $\mathcal{O}(10^{10})$ CPU-core-hours---i.e., a single analysis would keep $10^5$ CPU cores
busy for 30 years\footnote{Here we make the assumption that the algorithm can be parallelized perfectly.}---a restriction clearly prohibitive to performing any study.
Various state-of-the-art methods are employed to mitigate the high computational costs.
In the remainder of this section, we briefly present the main ideas behind these methods and give a conceptual introduction to how they are embedded in the approach we introduce in this article.

\subsection{Weighting and Smoothing}\label{sec:weighting}

The standard event-by-event MC weighting technique avoids repeated simulation and reconstruction of events every time a value of a nuisance parameter is changed.
This is possible, first, because the physics processes of neutrino production in the atmosphere (flux), their propagation involving flavor oscillation, and their detection and reconstruction are independent.
Each of these processes, therefore, can be treated separately.

For a process that has an a priori known parametric form (the parameter values of which are not necessarily known), the outcome of that process can be predicted by directly evaluating the parametrization at a set of input values.
In our case, both the neutrino flux prediction and flavor oscillations fall into this category.
The second category of processes are those that require MC simulation.
Predictions of the detection and reconstruction of neutrinos fall into this category because we do not have a complete characterization of the detector's response.

\begin{figure}[t]
    \centering
    \includegraphics[width=0.7\textwidth]{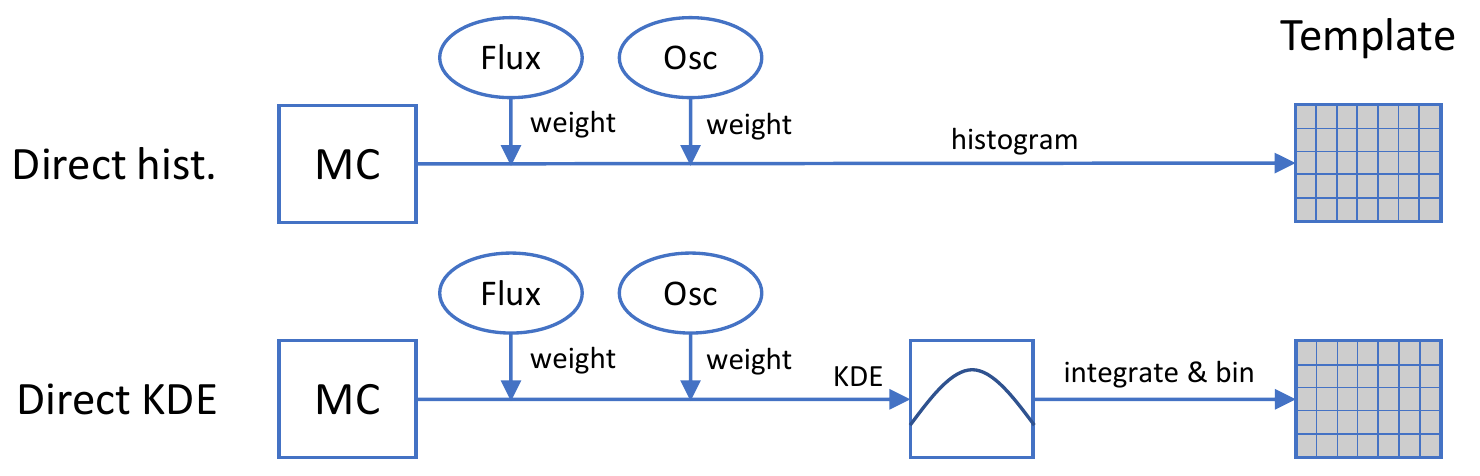}
    \caption{Operating principles of direct histogramming (top row) and direct KDE (bottom row), which both follow the same weighting scheme for MC events but arrive at the template differently, as explained in the text.}
    \label{fig:direc_modes}
\end{figure}

This leads to the standard event-by-event reweighting scheme, which estimates the expected final-level event counts due to all processes by simulating a set of MC neutrinos (capturing the effects of detection and reconstruction), assigning to each a weight derived from flux and oscillation calculations, and binning the events' weights in some set of observable dimensions, as illustrated in the top row of Figure~\ref{fig:direc_modes}.

In detail: Each MC neutrino---generated with a flavor $\beta$ and a set of true observables $\boldsymbol{\theta}^{\rm true}_\nu$---is assigned a posteriori the weight $w_\beta$ corresponding to the sum over the atmospheric fluxes $\Phi_\alpha(\boldsymbol{\theta}_{\rm flux};\boldsymbol{\theta}^{\rm true}_\nu)$ of all initial flavors $\alpha$ including the probabilities $P^{\rm osc}_{\alpha \to \beta}(\boldsymbol{\theta}_{\rm osc};\boldsymbol{\theta}^{\rm true}_\nu)$ to oscillate into a neutrino of the flavor $\beta$:
\[
w_\beta \propto \sum_\alpha \Phi_{\rm \alpha}(\boldsymbol{\theta}_{\rm flux};\boldsymbol{\theta}^{\rm true}_\nu) \times P^{\rm osc}_{\alpha \to \beta}(\boldsymbol{\theta}_{\rm osc};\boldsymbol{\theta}^{\rm true}_\nu).
\]
In the above, $\boldsymbol{\theta}_{\rm flux}$ and $\boldsymbol{\theta}_{\rm osc}$ are nuisance parameters affecting neutrino fluxes and oscillation probabilities. For a given realization of non-parametric nuisance parameters $\boldsymbol{\theta}_{\rm det}$ affecting the detector response, applying the detector response simulation (including event reconstruction, classification, etc.) to each incident neutrino results in a set of reconstructed observables $\boldsymbol{\theta}^{\rm reco}_\nu$, whose distribution can be compared to real data. In practice, techniques dealing with the discrete nature of detector nuisance parameters may be required. Here, however, we consider only a single realization of the detector parameters ($\boldsymbol{\theta}_{\rm det}$ fixed)---a simplification without any loss of the general applicability of the methods discussed.

Since the process of oscillation is decoupled from the detector simulation, only a single MC set is required to generate the templates for the different hypotheses under test (e.g., the two mass orderings); only the weights $w_\beta$ must be recomputed.
This eliminates statistical fluctuations between the otherwise disjoint MC samples.
However, even with a single MC set, an undersampling of the phase space of the model can result in a bias.

Binning the weights in (a relevant subset of) $\boldsymbol{\theta}^{\rm reco}_\nu$ corresponds to performing MC integration of the experiment's event distribution.
While the convergence rate of this approach does not depend on the dimensionality of the integral, errors of the estimates scale as $1/\sqrt{N}$, where $N$ is the number of MC events that fall in a bin.

As it is often infeasible to generate enough MC events to obtain sufficient accuracy in the MC integration process, smoothing of the final event distributions is a common practice. This, however, can be computationally slow and can introduce artificial features which may incorrectly reduce or enhance the signal. One such smoothing technique is kernel density estimation (KDE)~\cite{Cranmer:2000du}. Specifically, we apply adaptive bandwidth KDE directly to the weighted MC to compare a state-of-the-art version of this method to the methods we introduce in this paper in Section~\ref{sec:stage overview}. 
Here, a Gaussian kernel with a width calculated as described in~\cite{scott} is centered at each MC event's reconstruction information.
A weighted sum over the kernels of all events then delivers the smoothed distribution as shown in the bottom row of Figure~\ref{fig:direc_modes}, which will be compared to the distribution our method yields.\\

Shortcomings of the direct application of the two techniques discussed above---the first is the weighting method alone (labeled {\it direct histogramming}), while the second applies additional smoothing using adaptive kernel density estimates (labeled {\it direct KDE})---can be overcome using the {\it staged approach}.
Before providing an overview of the staged approach in Section~\ref{sec:stage overview}, we briefly introduce the key points of the example NMO analysis used to illustrate the benefits of the approach with respect to the standard techniques.

\section{NMO Analysis}
\label{sec:nmo}
The observation of neutrino oscillations and the demonstration of the neutrinos' non-zero masses~\cite{Fukuda:1998mi, Ahmad:2001an} represented a major step forward in the field of particle physics.
While current experimental techniques have not yet allowed for a direct measurement of the tiny masses, the magnitudes of their relative differences (mass splittings) are well known.

The ordering of these neutrino mass states (neutrino mass ordering, NMO) presents a difficult challenge. A powerful method to determine this ordering is the observation of matter effects on neutrinos. Owing to the high electron density of the Sun, observations of solar neutrinos have shown the second mass state to be heavier than the first~\cite{Aharmim:2011vm}. It  remains an open question, however, whether the third state is the most or least massive. The former scenario is referred to as the normal ordering (NO), while the second is called inverted ordering (IO). There is currently no experimental evidence decisively excluding either of the two scenarios~\cite{Capozzi:2018ubv,deSalas:2018bym,nufit4.0,Esteban:2018azc}.

The study of oscillations of atmospheric neutrinos provides a promising route toward a decisive measurement of the NMO~\cite{Patrignani:2016xqp,TheIceCube-Gen2:2016cap,PINGU-LOI,Adrian-Martinez:2016fdl}.
The path length (or {\it baseline}) varies between $\SI{20}{\kilo\metre}$ for vertically downward going and $\SI{12700}{\kilo\metre}$ for straight upward going atmospheric neutrinos, with the latter crossing the full diameter of the Earth.
With energies ranging from $\si{\mega\electronvolt}$ up to the $\si{\tera\electronvolt}$ scale, combinations of baselines and energies varying over several orders of magnitude are probed.
For the longest baseline, the very pronounced first oscillation maximum of muon neutrinos occurs at a neutrino energy of around $\SI{25}{\giga\electronvolt}$, followed by subsequent maxima at lower energies.

The electron neutrinos' coupling to electrons (coherent forward scattering) in the Earth creates an effective matter potential which leads to resonant behavior of the transition probabilities at energies around $\SI{5}{\giga\electronvolt}$, known as matter resonances~\cite{Wolfenstein:1978ue,Mikheev:1986wj,Petcov:1986qg}.
Furthermore, the Earth's specific density profile encountered by the neutrinos can also parametrically enhance their oscillations~\cite{Akhmedov:2005yj}. This enhancement with respect to oscillations proceeding in vacuum occurs for neutrinos if the NMO is normal, otherwise for anti-neutrinos.

\begin{figure}[t]
    \centering
    \includegraphics[scale=.45]{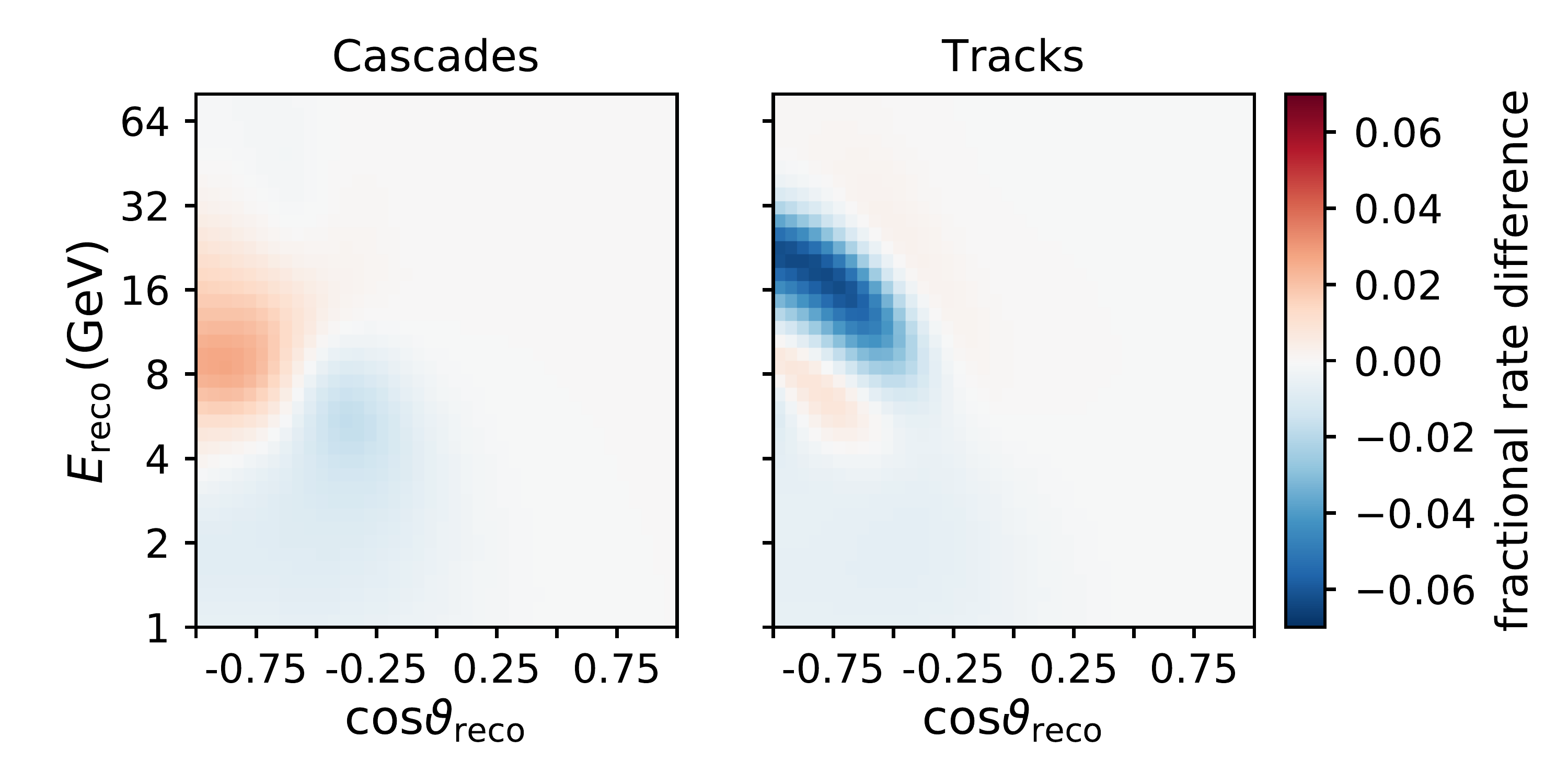}
    \caption{Expected fractional event rate difference between nominal NO and IO inputs (from Table~\ref{table:params}) for the toy model. Cascades are shown on the left, tracks on the right.}
    \label{fig:nmo-signature}
\end{figure}

The NMO measurement potential of {\vlvnt}s is based on this asymmetry. Two major aspects are obstructive, however. The first is the inability of {\vlvnt}s to differentiate between neutrinos and anti-neutrinos. This reduces the effect to the respective difference in atmospheric fluxes and interaction cross sections. Energy and directional resolutions of the experiment present the second hurdle. Both are typically prohibitive to resolving the fast variations of the oscillation pattern at the relevant energies. As a consequence, the observable effect is reduced to at most a few percent over the relevant energy and zenith range (see Figure~\ref{fig:nmo-signature}), requiring neutrino telescopes with effective masses on the order of megatons to achieve sufficient event statistics.

Proponents of various {\vlvnt}s in ice and water have performed studies confirming this idea, finding that a $>3\sigma$ (median) sensitivity to the NMO can be achieved within five years of exposure time even in less favorable regions of the neutrino oscillation parameter space~\cite{TheIceCube-Gen2:2016cap,Adrian-Martinez:2016fdl,Abe:2011ts}.

As the oscillation probabilities directly depend on neutrino energy {\Etrue}, oscillation baseline ($\propto\CZtrue$), and flavor, we split our data into bins of $\log_{10}${\Ereco}, {\CZreco}, and event class\footnote{The use of the subscript ``true" is used to specify the true variables of the neutrinos and to distinguish these from the reconstructed variables, denoted with a subscript ``reco", which will be introduced in Section~\ref{subsec:stages}.}. It is important to choose a binning fine enough to resolve the NMO signature, while coarse enough to retain a sufficient amount of MC statistics per bin, as motivated in Section~\ref{sec:prob}.
We have found the division into $(40 \times 40 \times 2)$ bins to be suitable, covering a range of {\Ereco} from \SIrange{1}{80}{\giga\electronvolt}, the whole sky ($\CZreco$ from $-1$ to $1$), and the two event classes of {\it cascades} and {\it tracks}.
Using this binning, for our toy detector introduced in~\ref{sec:toy}, Figure~\ref{fig:nmo-signature} shows the expected fractional event rate difference $(R_{\rm NO}-R_{\rm IO})/R_{\rm NO}$, where $R_{\rm NO(IO)}$ is the expected event rate for true NO (IO), based on the two sets of nominal model parameter values given in Table~\ref{table:params}.

\begin{figure}[ht]
    \centering
    \includegraphics[width=.5\textwidth]{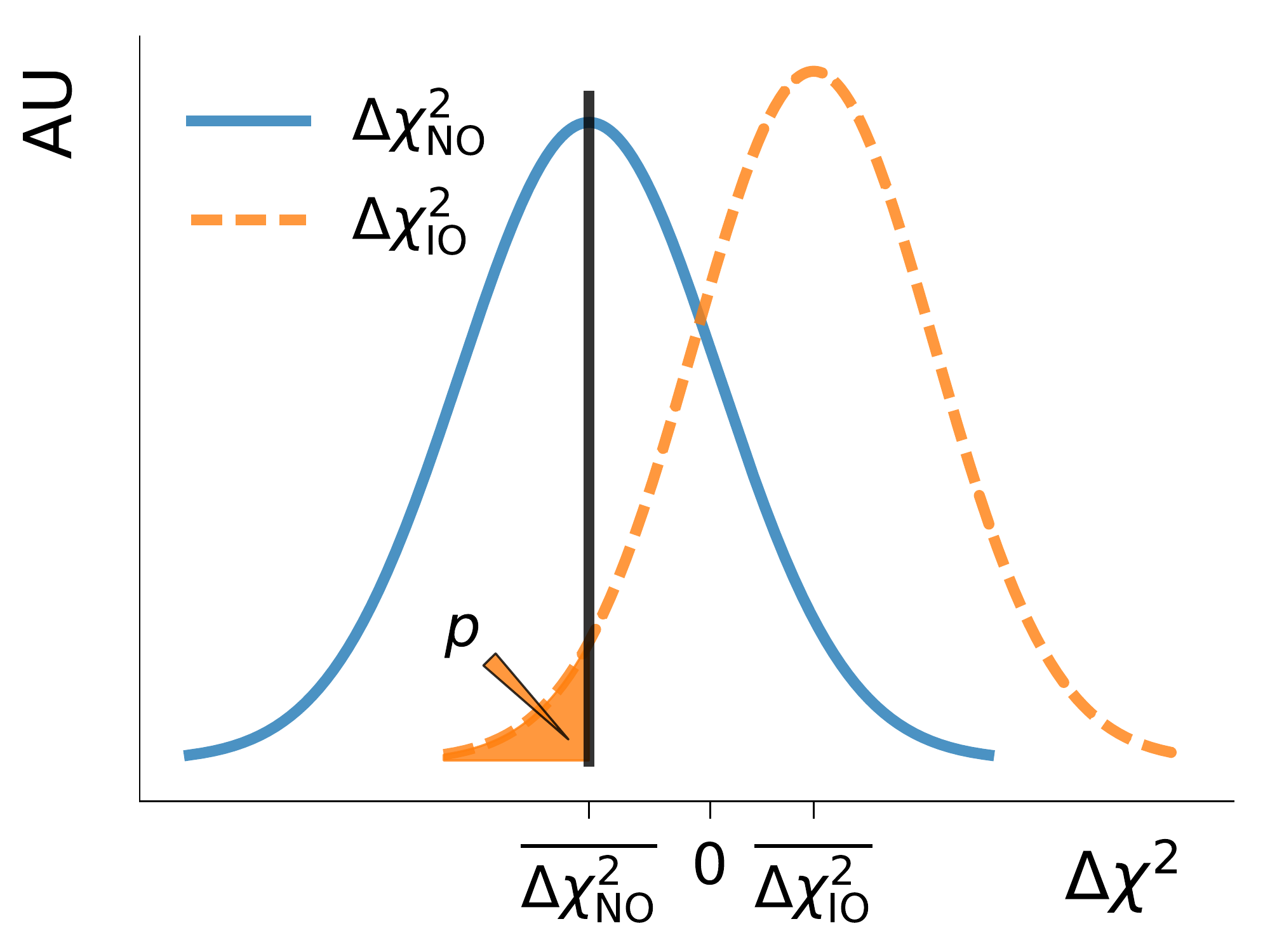}
    \caption{Example distributions of Equation~(\ref{eq: delta chisquare}). The distribution on the left (solid line) represents the case of NO pseudo-data, while the distribution on the right (dashed) is obtained when the pseudo-data is taken from the IO. Here, $1-p$ corresponds to the confidence level at which the IO is correctly rejected with a probability of 50\%.}
    \label{fig: ts dist}
\end{figure}

As the most powerful test statistic for distinguishing two simple hypotheses \cite{Neyman-Pearson}, the \emph{logarithm of the likelihood ratio}
\begin{equation}
\mathcal{T} = -2\ln\left(\frac
{\max\limits_{\boldsymbol{\theta} \in \rm{NO}}L(\mathbf{n} | \boldsymbol{\mu}(\boldsymbol{\theta}))}
{\max\limits_{\boldsymbol{\theta} \in \rm{IO}}L(\mathbf{n} | \boldsymbol{\mu}(\boldsymbol{\theta}))}\right)\text{ .}
\label{eq: llr}
\end{equation}
is also useful in assessing the ability of an experiment to discriminate between the two (composite) NMO hypotheses at a given confidence level.
It is representative of the degree at which observing the data $\mathbf{n}$ under the NO hypothesis is favored over observing it under the alternate IO hypothesis. The observed spectrum at the detector, $\mathbf{n}$, however, is a convolution of the atmospheric neutrino flux, the effects of neutrino oscillations that bear the NMO signature, the neutrino interaction and detection processes, and the event reconstruction and classification procedure. Each one of these effects is accompanied by systematic uncertainties. As their impact on the predicted spectrum $\boldsymbol{\mu}$ is modeled, the systematic uncertainties directly feed in to the likelihood $L$ of the observation.

For this study, we limit ourselves to a simplified treatment using $\chi^2$ statistics and the {\it Asimov} dataset. In this approach, the projected median sensitivity is calculated from the average experimental outcomes under the two possible NMO hypotheses, as opposed to performing extensive ensemble tests with randomly fluctuated pseudo-experiments.
The log-likelihood expression is a simple $\chi^2$, and Equation~(\ref{eq: llr}) can be rewritten as the difference
\begin{equation}
    \Delta\chi^2 = \chi^2_{\rm NO} - \chi^2_{\rm IO}\text{ .}
    \label{eq: delta chisquare}
\end{equation}
Here, $\chi^2_{\rm NO}$ is the minimum $\chi^2$ between model predictions and data, with all nuisance parameters profiled out using NO priors ($\chi^2_{\rm IO}$ follows analogously).

An illustration of example distributions of (\ref{eq: delta chisquare}) for the two different NMO hypotheses is shown in Figure~\ref{fig: ts dist}. The goal is to obtain a p-value $p$ which quantifies the statistical compatibility between the hypothesis that is tested and the one assumed to be true. In the ensemble approach, the two distributions would need to be built up by fitting pseudo-experiments. In the Asimov approach, however, certain assumptions about the distribution of (\ref{eq: delta chisquare}) allow adopting the expression $\sqrt{|\overline {\Delta \chi^2}|}$ as a sensitivity proxy~\cite{Blennow:2014fk}, determining the significance at which the wrong ordering can be excluded without the need for pseudo-experiments.

For the profiling of the nuisance parameters (any free model parameters), a numerical algorithm minimizes the $\chi^2$ metric. Whenever external constraints are applied to such parameters, we add those to the $\chi^2$ value as penalty terms (priors). While the presence of these penalty terms is meant to illustrate a typical approach to problems of this sort, their sizes do not follow any precise physical motivation. Table~\ref{table:params} gives an overview of all used model parameters, their nominal values for NO and IO, and priors (where applied). 

\begin{table}[tb]
\centering
\label{my-label}
\begin{tabular}{l|c|c|c}
\multirow{2}{*}{Parameter}     & \multicolumn{2}{c|}{Nominal value} & \multirow{2}{*}{Prior}\\\cline{2-3}
          & \multicolumn{1}{c|}{NO}              & IO               &      \\ \hline
{\nue/\numu} flux ratio            & 1.0 & 1.0          & $\pm 0.03$ \\
$\nu/\bar{\nu}$ flux ratio       & 1.0 & 1.0           & $\pm 0.1$  \\
Spectral index shift             & 0.0 & 0.0           & $\pm 0.1$  \\
Energy scale                     & 1.0 & 1.0           & $\pm 0.1$  \\
Overall normalization            & 1.0 & 1.0           & $\pm 0.1$  \\
$\theta_{13}$ ($^{\circ}$)             & 8.5 & 8.5           & $\pm 0.2$~\cite{Gonzalez-Garcia:2014bfa,nufit2.0}  \\
$\theta_{23}$ ($^{\circ}$)             & 42.3            & 49.5             & non-Gaussian~\cite{Gonzalez-Garcia:2014bfa,nufit2.0} \\
$\Delta m_{31}^2$ (${\rm eV^2}$) & 0.00246         & -0.00237         & $\pm \SI{4.75e-5}{}$~\cite{Gonzalez-Garcia:2014bfa,nufit2.0}

\end{tabular}
\caption{Summary of model parameters in the example NMO analysis, including their nominal values for the two NMO hypotheses and Gaussian $\pm 1\sigma$ bounds used as external constraints (priors). The first three parameters are applied to atmospheric neutrino flux predictions from~\cite{Honda:2015fha}, following the procedure laid out in Section~\ref{sec:flux}. The values for the three oscillation parameters are based on a recent global fit~\cite{Gonzalez-Garcia:2014bfa,nufit2.0}.}
\label{table:params}
\end{table}

\section{Overview of the Staged Approach}
\label{sec:stage overview}
\label{sec:stages}

\begin{figure}[tb]
  \begin{center}
    \includegraphics[width=14cm]{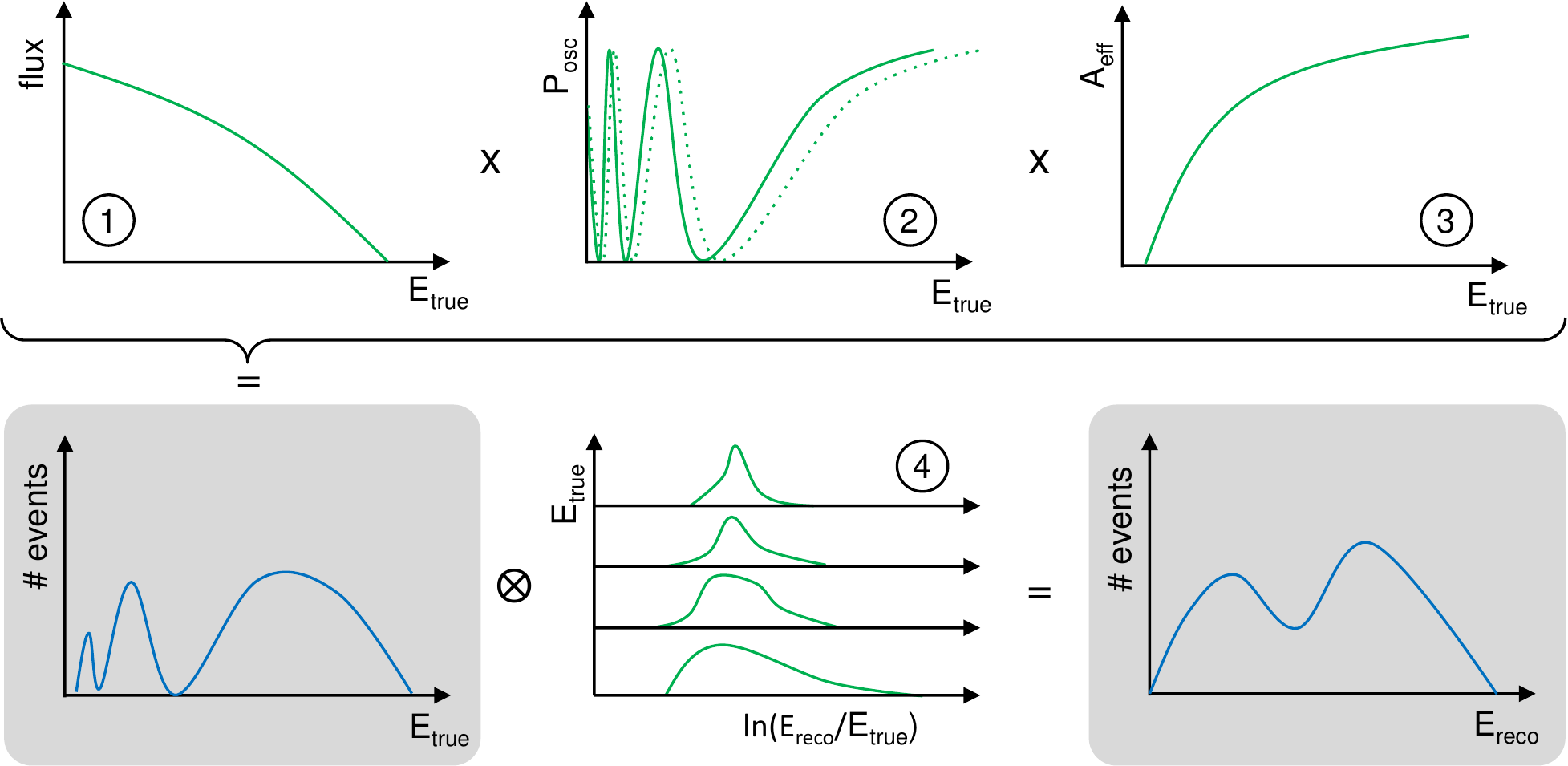}
  \end{center}
  \caption{
  Illustration of the staged approach for obtaining event templates, here for simplicity using a characterization in one dimension (energy) only.
  Steps 1, 2, and 3 are in true energy ({\Etrue}); the product of these yields the expected event distribution (lower left).
  Smearing this spectrum with energy-dependent energy resolution functions (step 4) gives the reconstructed event rate spectrum (lower right).
  Note that the dotted green line in step 2 shows a hypothetical change of oscillation parameters, affecting only stage 2.
  Smoothing can now directly be applied to the distributions in steps 3 and 4, instead of the fully weighted MC as in the direct KDE method.}
  
  \label{fig:pisa_comic}
\end{figure}

The method to obtain templates we describe in this article is divided into four independent parts, referred to as {\it stages}.
The four stages (flux, oscillation, detection, and reconstruction) and how they are used to obtain event templates are summarized in this section, while more technical descriptions of each stage follow in Section~\ref{sec:implementation}.

\subsection{Stages}
\label{subsec:stages}
Templates for our example case of an NMO analysis using a  {\vlvnt} are produced efficiently and accurately using the following four stages.

Each stage represents a collection of related physical effects. Beginning with the flux computed by the initial stage, each subsequent stage applies a transformation to the output of the previous stage.

\begin{enumerate}
    \item {\bf Flux}
The expected unoscillated atmospheric neutrino fluxes are taken from an external model~\cite{Honda:2015fha}.
Flux values from this model are provided in the form of tables with discrete steps in both neutrino energy, {\Etrue}, and direction, here the cosine of the zenith angle, {\CZtrue}. Therefore, an interpolation must be performed for values between those tabulated.
Crucially, these tables give the integrated flux across the bins, which does not necessarily coincide with the flux value at the bin center.
Accordingly, we use an integral-preserving (IP) interpolation.
In general, atmospheric flux models
require external inputs including primary cosmic ray measurements, atmospheric density models, and hadronic interaction measurements.
Many associated uncertainties are known~\cite{Barr:2006it, Evans:2016obt} and need to be included as nuisance parameters in an analysis.

    \item {\bf Oscillation}
Flavor oscillations of neutrinos traversing the Earth modify the flavor content of the original flux in a manner that depends on the energies and path lengths (derived from the direction) of the neutrinos.
Additional intrinsic neutrino properties determine
the standard flavor oscillation probabilities: three mixing angles and two independent mass-squared splittings, as well as a possible non-zero CP-violating phase.
In addition, matter effects induce modifications in the flavor transition probabilities compared to vacuum~\cite{Mikheev:1986wj,Wolfenstein:1978ue,PhysRevD.22.2718}, which makes up the basis of the NMO measurement capability of {\vlvnt}s.
In~\cite[]{PhysRevD.22.2718}, the authors present an analytical expression for the neutrino flavor transition amplitude in a layer of uniform-density matter, which in turn was later implemented in, for example, the \texttt{Prob3++} software~\cite[]{prob3}.
Here, the Earth density profile~\cite[]{Dziewonski1981297} is approximated by a finite number of homogeneous layers and the total transition amplitude is represented by a matrix product of the amplitudes in the individual layers.
The main challenge for this stage, which in contrast to the other stages does not require any MC simulation, is to keep the burden of these computationally expensive calculations to a minimum, while retaining sufficient accuracy in the modeling of the neutrinos' propagation.

    \item {\bf Detection}
The number of observed events is determined by the (oscillated) flux as well as a quantity known as the {\it effective area} (alternatively, the effective mass)\footnote{In contrast, high energy physics experiments often calculate an acceptance instead, which is also based on simulation.}.
This incorporates the probability that a given neutrino interacts within the detector, is detected, and passes the given data selection criteria.
We obtain the eight effective areas ($\nu_{e,\mu,\tau}$ \& $\bar{\nu}_{e,\mu,\tau}$ charged current (CC) and $\nu$ \& $\bar{\nu}$ neutral current (NC) interactions) from simulated MC events that are run through the same selection criteria as the real data.
In general, each of these effective areas will depend on the energy and arrival direction of the neutrinos.
Depending on the detector geometry, certain symmetries can be exploited to reduce the number of parameters on which the effective areas depend.
Here we assume azimuthal symmetry and therefore only extract effective areas as a function of {\Etrue} and {\CZtrue}.

    \item {\bf Reconstruction}
The process referred to as \emph{reconstruction} translates the raw signals recorded by a detector into estimates of the physical properties of events.
Uncertainties in these estimates manifest as statistical fluctuations, with respect to the true properties, which can be described by probability density functions we refer to as \emph{resolution functions}.
We estimate the resolution functions from the same MC events as used in the detection stage, for which we know the true energy, zenith angle, and interaction type on an event-by-event level.
The reconstruction stage uses these estimated resolution functions to build smearing kernels (ensembles of resolution functions) that map the event rates from the space of true variables into the space of reconstructed observables. Additionally---since most {\vlvnt}s cannot exactly distinguish the different neutrino flavors and interaction types---the events are classified by their signature in the detector.
Here, event classes are {\it tracks} and {\it cascades}, based on whether the event seems to contain the expected signature of a starting muon track.
This process will separate {\numucc} and {\numubarcc} interactions from all others, albeit with limited efficiency and purity.
For the example NMO analysis, three observables are needed: the primary neutrino's reconstructed energy (\Ereco), zenith angle ($\vartheta_{\rm{reco}}$), and event classification.

\end{enumerate}

\begin{figure}[tb]
  \begin{center}
    \includegraphics[width=14cm]{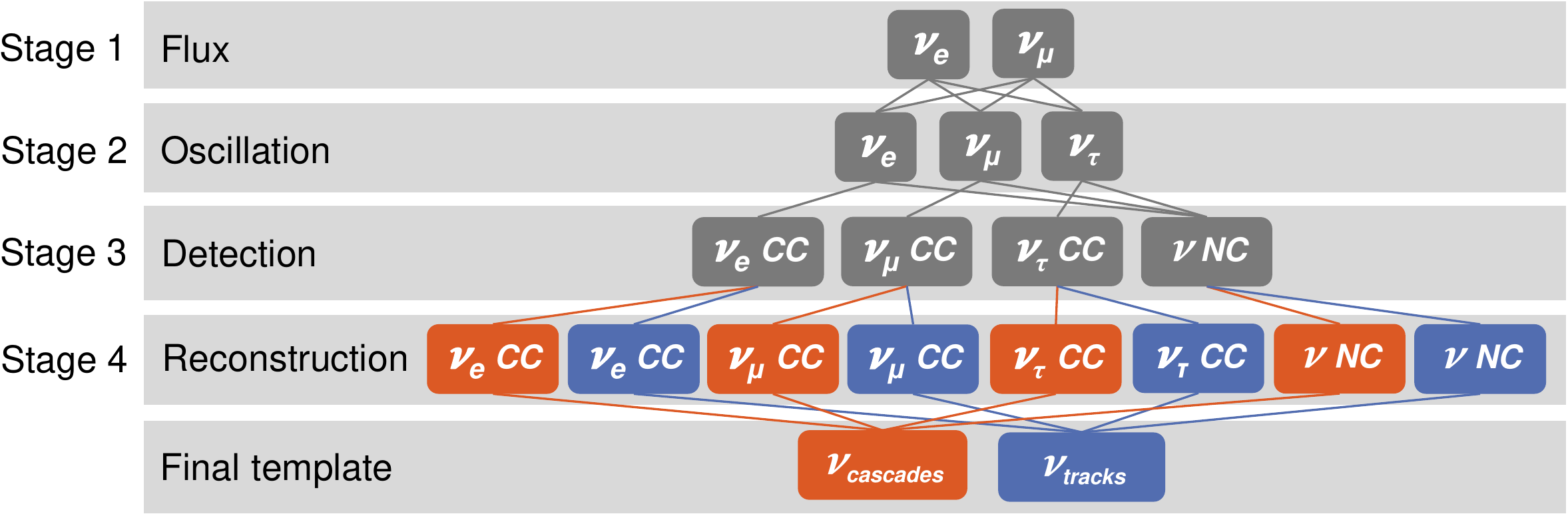}
  \end{center}
  \caption{
  Flow of neutrino flavors and interaction types through the stages, here shown for neutrinos only (with an analogue counterpart for anti-neutrinos).
  Neutral current events of all flavors are indistinguishable and can therefore be conveniently added together.
  The reconstruction stage not only maps from ($\Etrue \times \CZtrue$)-space to ($\Ereco \times \CZreco$)-space, but also classifies the events into the cascade and track categories, indicated by the orange and blue color, respectively.
  }
  \label{fig:pisa_tree}
\end{figure}

Note that there is no
universal prescription for identifying the set of stages appropriate for
any given physics analysis or detector.  Instead, stages are chosen to
exploit valid simplifications for the task at hand.  For example,
atmospheric neutrino flux and oscillation calculations depend on readily
available tabulated spectra and analytic formulae, respectively.  Cosmic
ray observatories or high energy particle colliders, by contrast, might
require complex stages to describe particle showers, which in turn might
depend on high-dimensional, analysis-specific tables.  Any physics
scenario resulting in multi-particle final states adds further
complexity.

In essence, the specific problem and analysis at hand determine to which extent MC sampling is necessary and whether the staged approach is applicable. If the latter is indeed the case, care must be taken concerning the choice of appropriate stages and their specific implementations.
In the remainder of this article, we study in detail the staged approach we have found particularly effective for an NMO analysis using a {\vlvnt}.

\subsection{Template Generation}
\label{subsec:template gen}

In order to produce the final-level event templates that are ultimately compared to the data, the four stages are combined as depicted in Figure \ref{fig:pisa_comic}: integration of the product of the first three stages (flux, oscillation probability, and effective area) over {\Etrue} and {\CZtrue} yields the event rate in the space of true variables.
The event rate in the space of reconstructed observables is then obtained by a convolution of the true event rate with the reconstruction resolution functions.
Finally, multiplication by detector exposure time results in an event {\it count}, which can be compared directly to observed data or different templates\footnote{While not shown here, it is possible to extend the model with more parameters or stages to describe additional effects, such as the modeling of systematic uncertainties.}.

Throughout the stages, different combinations of neutrino flavor and interaction type (channels) need to be treated separately, as depicted in Figure~\ref{fig:pisa_tree}.
Starting with the atmospheric flux, the neutrinos can undergo flavor change via oscillation.
Since {\nutau} production in the atmosphere is expected to be negligible
at the energies relevant here, this flavor only appears through oscillation \cite{Bulmahn:2010pg}.    
The detection rate varies between CC and NC interactions \cite{Patrignani:2016xqp}.
Finally, after applying the reconstruction resolutions and event classification, event counts are summed to get the final-level templates for events classified as tracks and cascades separately.
Where not mentioned explicitly, the same treatment is also applied to anti-neutrinos. The final templates are the sum over both, neutrinos and anti-neutrinos.

Since the transformations computed by individual stages are independent of one another, a parameter change affecting one stage does not affect the transformations used by the other three stages, and in particular not the result of the previous stages.
Therefore, we include caching functionality that reduces the overall computational expense when a number of successive templates are retrieved while changing one parameter at a time.

The transformations performed by the individual stages are dependent on the neutrino's energy and zenith angle, and therefore must be computed and applied differentially.
All stages are evaluated on a grid of points distributed over {\Etrue} and {\CZtrue}, with the final templates output in {\Ereco}, {\CZreco}, and event class. Points in energy are logarithmically spaced in the domain \SIrange{1}{80}{\giga\electronvolt} while points in cosine-zenith are linearly spaced between $-1$~and~$1$.
The number of bins
in each stage (for input, transformation, and output) is adjusted to reduce numerical integration errors and to avoid smearing out the physical effects under study.
At the same time, this number should be kept as small as possible to reduce the computational load.
An overview of the binning scheme we have employed, suitably mediating between these two effects, is given in Table~\ref{table:stage binning}.

\begin{table}[tb]
\centering
\begin{tabular}{l|ccc}
Stage           & Transformation                                & Output                  \\
\hline       
Flux            & -                                             & 400 {\Etrue} $\times$ 400 {\CZtrue} \\
Oscillation     & $400 \times 400$                              & 400 {\Etrue} $\times$ 400 {\CZtrue} \\
Detection       & $400 \times 400$                              & 200 {\Etrue} $\times$ 200 {\CZtrue} \\
Reconstruction  & $200 \times 200 \times 40 \times 40 \times 2$ &  40 {\Ereco} $\times$  40 {\CZreco} $\times$ 2 classes
\end{tabular}
\caption{
    Gridpoints chosen for the staged approach in this work.
    The output of one stage is the input to the next stage, and the result of the detection transformation is downsampled from (400$\times$400) to (200$\times$200) by summing non-overlapping sets of 2$\times$2 adjacent points.
    Outputs of flux, oscillation, and detection are in the domain {\Etrue}~$\in$~(1,~80)~GeV and {\CZtrue}~$\in$~(-1,~1) while the output of reconstruction is in the domain {\Ereco}~$\in$~(1,~80)~GeV, {\CZreco}~$\in$~(-1,~1), and class~$\in$~\{track,~cascade\}.
    Within their respective domains, points in energy are logarithmically spaced while points in cosine-zenith are linearly spaced.
}
\label{table:stage binning}
\end{table}

The fundamental motivation for splitting up the process of template generation into a sequence of stages is that smoothing methods can be chosen for each stage that accurately reflect their unique physics, which in our example analysis apply to the detection and reconstruction stages.
This approach reduces the required MC statistics with no loss of detail in the flux and flavor oscillation modeling.
In contrast, smoothing events at the final level, as the traditional direct KDE does, acts on a convolution of effects, including the rapidly-varying behavior in the underlying oscillation physics.
As will be shown later in this article, this difference is key to achieving higher precision with the staged approach compared to our reference methods.

For the staged approach, we emphasize that our choice of smoothing techniques is not unique. The specific techniques we employ are motivated by the typical shapes of the distributions characterized and have been found to be reliable and robust at modest computational costs. They should thus be seen as effective but non-exclusive solutions to problems of the kind discussed in this article.

\begin{figure}[t]
    \centering
    \includegraphics[width=0.7\textwidth]{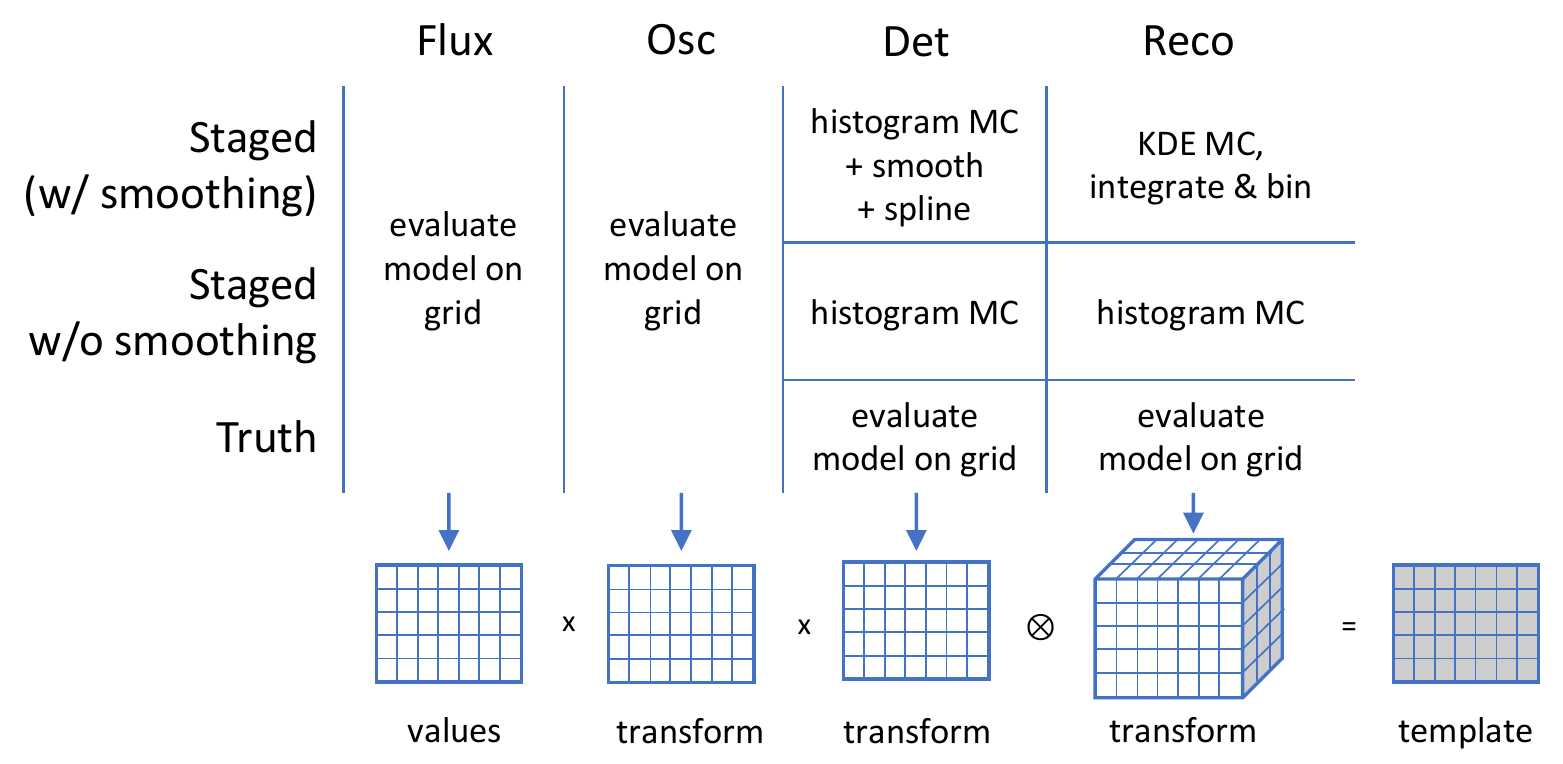}
    \caption{
    Operating principles of the different staged approach modes, which differ in how we generate the transformations of the last two stages. The staged approach without smoothing is employed for validation purposes in Section~\ref{subsec:sampling grid}. See text for details. }
    
    \label{fig:stage_modes}
\end{figure}

Note that, in addition to the MC-based calculation of the transformations provided by the detection and reconstruction stages, we have implemented the option to produce transformations using the parametric functions of the toy model defined in~\ref{sec:toy}. The template produced in this way is what we refer to as ``truth''.

All MC events we use with the staged approach are samples of the unbinned distributions of the toy model and are shared between the detection and reconstruction stages. For each combination of neutrino type and interaction type (for example $\nu_e$ CC, $\bar{\nu}_\mu$ NC), we draw an identical number of events.
This number, one twelfth of the total number of events constituting a given random sample of the toy model, is referred to as the \emph{sample size}. A given sample is used together with the event-by-event MC weighting technique to generate templates for all possible values of $\boldsymbol{\theta}$, that is, to calculate the associated expected counts in all bins of each final-level template.

A complete overview of the different operation modes of the staged approach is given in Figure~\ref{fig:stage_modes}, which highlights the stages at which these differ in the template generation process.

\section{Technical Implementation of Stages}
\label{sec:implementation}
The stages within our approach, as summarized in Section~\ref{sec:stage overview} and illustrated in Figure~\ref{fig:pisa_comic} are subject to different technical and computational challenges due to the physics effects captured by each one.
In this section we examine specific implementation details which highlight how each stage balances performance and precision requirements---even in the presence of low MC statistics.

Therefore, we include caching functionality that reduces the overall computational expense when a number of subsequent templates are retrieved while changing one parameter at a time.

\subsection{Flux}
\label{sec:flux}

In order to preserve the integral of a tabulated set of data, a spline is fit to the {\it integral} of the data rather than to the values themselves.
Interpolated values in the initial space are then found by evaluating the derivative of these splines.
We refer to this method as integral-preserving (IP) interpolation.

For the NMO example analysis, the tabulated data of interest are the atmospheric neutrino flux predictions from~\cite{Honda:2015fha} provided as a function of both {\Etrue} and {\CZtrue}.
To simplify the problem, the integration\footnote{Here, a cumulative sum of the bin values multiplied by the respective bin width.} is performed along one dimension at a time.

Consider the case with fluxes tabulated at $M \times N$ points in ({\Etrue}, {\CZtrue}).
To retrieve the flux at an arbitrary ({\Etrue}, {\CZtrue}) point, say $(x, y)$, first one spline of the integrated flux as a function of {\CZtrue} is created for each of the $M$ {\Etrue} locations.
The derivative of each of these splines is evaluated at $y$, yielding $M$ flux values.
The integral of these values is then interpolated with a spline, and finally this spline's derivative is evaluated at $x$.
This concept generalizes to higher dimensions, but can quickly become computationally intensive as the number of splines grows.
While the splines used in the provided example are of one-dimensional cubic type, other spline variants or interpolation techniques can be used, as long as these allow for differentiation.
For the example analysis of this article, IP interpolation is approximately an order of magnitude slower than two-dimensional cubic spline interpolation.

\begin{figure}[tb]
  \begin{center}
    \includegraphics[width=16cm]{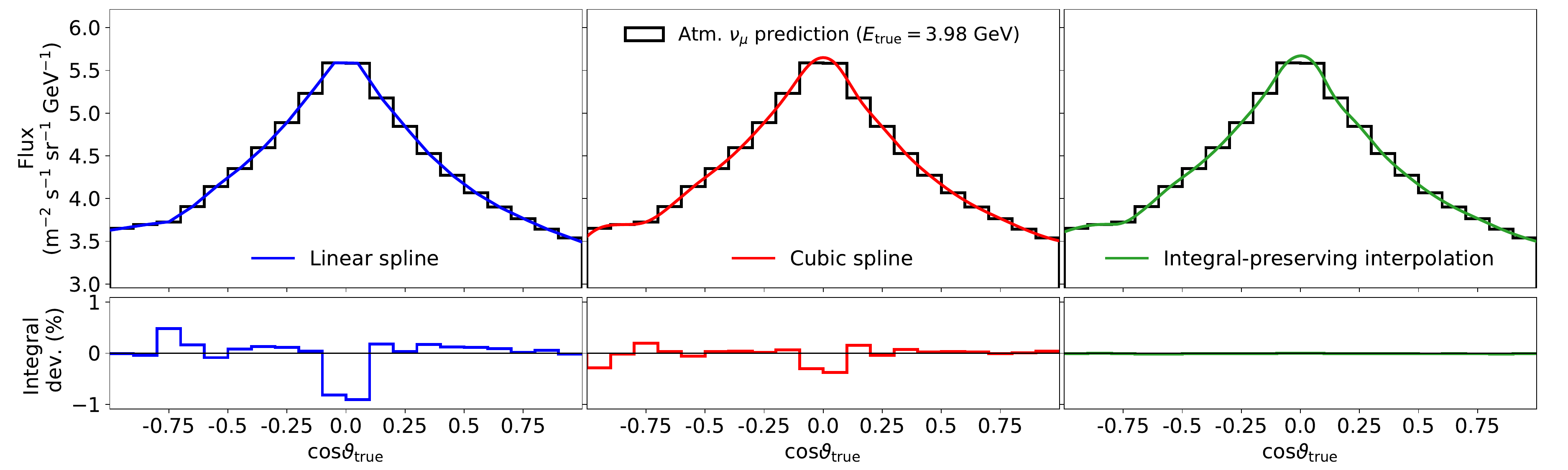}
  \end{center}
  \vspace{-15pt}
  \caption{The top part of the figure shows three different interpolation methods applied to the same set of data points from~\cite{Honda:2015fha} while the bottom portion shows the fractional deviation of the integral (= area under the curve) from these three methods. The deviations from the integral-preserving method presented in this paper have a maximum \(\sim\)0.02\%.}
  \label{fig:ToyInterpolationProblem}
\end{figure}

The IP method improves upon standard interpolation techniques in that it correctly models the turnover of the flux at the horizon $\left( \CZtrue=0 \right)$ and the behavior in the most upgoing and downgoing regions $\left( \CZtrue \sim \pm1 \right)$.
This can be seen in Figure~\ref{fig:ToyInterpolationProblem}, which compares the results of IP to linear and cubic spline interpolation.

For the tables used in this article's example analysis, IP interpolation preserves the integral to better than 0.5\% over the complete ({\Etrue}, {\CZtrue})-space.
More detailed information on the IP method can be found in \cite{Wren:2018}.

\subsection{Oscillation}
\label{sec:osc}

The oscillation library that we employ is an extension of the code described in~\cite[]{Calland:2013vaa}, where the authors ported some of the core functions of \texttt{Prob3++} to a graphics processing unit (GPU) via the CUDA C API \cite{Nickolls2008ScalablePP}---an application programming interface to perform general purpose computations on GPUs.
We then added back in the ability to handle an arbitrary number of constant density layers of matter, allowing for highly parallel calculations of three-flavor oscillation probabilities of neutrinos that encounter a realistic radial Earth density profile, with fine-grained control over its characteristics.
We implemented the oscillation calculations with floating point precision selectable to either single (32 bits, or FP32) or double (64 bits, or FP64) precision.
With our code run in double precision with \texttt{Prob3++}, evaluated on a 100$\times$100 grid of neutrino energies {\Etrue} ranging from \SIrange{1}{80}{\giga\electronvolt} and {\CZtrue} values spanning the region between $-1$ and $0$,
our GPU and CPU implementations of the \texttt{Prob3++} code produce consistent results to the level of $10^{-14}$ or less.
These differences are due to differing hardware implementations of the same mathematical operations.
Switching from double to single precision on the GPU, we find that the magnitudes of the differences stay below about $10^{-5}$ for all oscillation channels.
Single precision is desirable from a performance point of view, since most GPUs comprise a larger number of single precision than double precision arithmetic units, and these extra units can be exploited by the parallelism in our code.

\begin{figure}[tb]
    \centering
    \includegraphics[width=.8\textwidth]{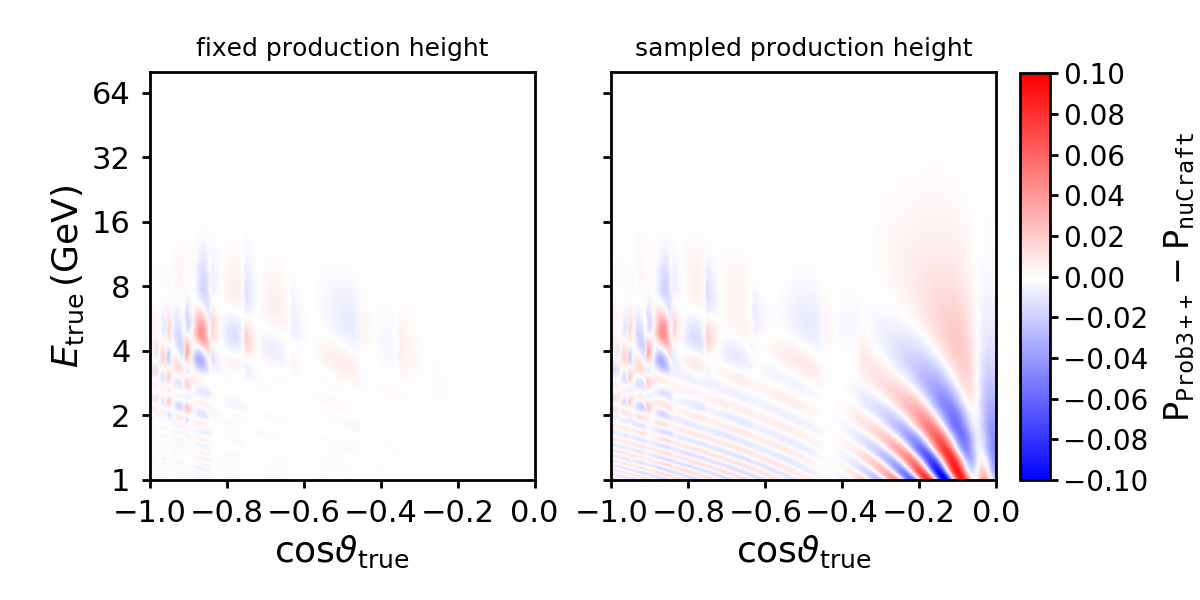}
    \caption{
    Deviation of {\numu} survival probabilities computed with \texttt{Prob3++} compared to \texttt{nuCraft}.
    The left panel uses a fixed production height of $\SI{20}{\kilo\metre}$ for both codes and twelve constant-density layers for \texttt{Prob3++}. In the right panel the values from \texttt{nuCraft} are the average probabilities for a range of neutrino production heights across the atmosphere.
    }
   \label{fig:nucraft vs prob3++}
\end{figure}

To evaluate the effects of an approximated Earth density profile using a limited number of constant density layers and a constant atmospheric production height---both approximations that our code makes---we compare the oscillation probabilities from our implementation of \texttt{Prob3++} against a reference model. The latter is calculated by \texttt{nuCraft}~\cite[]{Wallraff:2014qka}, which is written in Python and solves the Schr\"odinger equation numerically. The \texttt{nuCraft} library also supports a realistic variation of the oscillation baselines according to the distribution of atmospheric neutrino production heights described in~\cite[]{PhysRevD.57.1977} and uses an interpolated radial density profile of the Earth.

To this effect, we first fix the atmospheric neutrino production height to $h_0=\SI{20}{\kilo\metre}$ for both codes, and quantify the deviations arising from the coarser Earth model by calculating the {\numu} survival probability residuals on a fine grid in cosine zenith and energy.
When approximating the Earth's density profile with only four layers (one for each of the upper and lower mantle, and the outer and inner core), differences of up to $\SI{15}{\%}$ to the output of \texttt{nuCraft} are seen. These differences decrease to below $\SI{5}{\%}$ when using 12 density layers (see left panel of Figure~\ref{fig:nucraft vs prob3++}). Using an even more detailed model with 59 layers results in differences smaller than $\SI{0.3}{\%}$ across the whole two-dimensional spectrum.

Comparing the 12-layer \texttt{Prob3++} probabilities to those obtained under the assumption of a more realistic distributed atmospheric production height in \texttt{nuCraft} highlights further discrepancies between the outputs of the two codes (see right panel of Figure~\ref{fig:nucraft vs prob3++}). However, the largest differences ($\sim \pm \SI{10}{\%}$) appear for near horizontal trajectories, while the residuals for $\cos{\vartheta}\lesssim-0.4$ remain roughly unchanged.

Since precise modeling of both the Earth's density profile and the atmospheric neutrino production heights come at a significant additional computational cost, depending on the analysis in question it might be desirable (and justifiable) to neglect one or both of these effects.
In our example NMO analysis we find that it is sufficient to use the 12-layer model and a fixed production height.
Both approximations have very little impact on the final spectra---mainly due to detector resolution effects---and since they systematically affect both NMO realizations in an almost identical manner, their effects leave the measurement comparing the two mass orderings largely unaffected. 
Moreover, while the atmospheric flux peaks in horizontal direction (seen, for example, in Figure~\ref{fig:aeff_vs_stats}), negligible matter effects for the corresponding trajectories result in very little intrinsic sensitivity of this part of the spectrum to the NMO.

\subsection{Detection}
\label{sec:aeff}

As a reminder, the effective areas are quantities used to translate an incoming flux to the event rates in the detector.
These effective areas are calculated from a limited number of MC events, hence they can suffer from statistical fluctuations which can be a non-negligible contribution to the total uncertainty of the final physics result.
At the same time, effective areas are typically well-behaved quantities in energy and zenith angle (under some realistic assumptions, e.g., that no discontinuous selection cuts are applied and no gaps exist in the detector acceptance).
Therefore, smoothing techniques can be applied to alleviate the unwanted uncertainty contributions from statistical fluctuations.

For charged current interactions, we compute the effective area separately for each neutrino flavor.
In contrast, we do not distinguish between flavors for neutral current (NC) interactions, since their cross sections are identical.
Neutrinos and anti-neutrinos are handled independently, accounting for a total of eight independent effective area functions.
For convenience we include the multiplication by detector exposure time ($t_{\rm exp}$) in the same step, which means that this stage outputs event counts ($N_{\rm events}$) instead of rates
\begin{equation}
\label{eq:aeff}
    N_{\rm events} = \Phi [{\rm m^{-2}s^{-1}}] \cdot A_{\rm eff} [{\rm m^2}] \cdot t_{\rm exp}[{}\rm s]\,,
\end{equation}
for some input flux ($\Phi$).

In our staged approach we first evaluate the effective areas on a fine grid in ({\Etrue}, {\CZtrue}) using the MC events via MC integration, where, when generating events, the sampling is chosen to provide a relatively uniform coverage across all grid points. For our example case study, we use a uniform sampling across \CZtrue\ and a power law spectrum for the energies $\propto\Etrue^{-1}$ to closely follow actual IceCube oscillation analyses. (Note that an optimization of the sampling choices would benefit both the staged approach and the reference methods.)
Still, for small sample sizes, some grid points may have no associated events, leading to gaps in the distribution. We remove these by applying a simple Gaussian smearing along the two-dimensional grid. 
In a second step, cubic splines are employed to perform smoothing. Here, first, splines are created along the {\Etrue} dimension individually for every {\CZtrue} bin, and evaluated to obtain new values for every grid point. Then, this splining procedure is repeated along the {\CZtrue} dimension.

\begin{figure}[tb]
  \begin{minipage}{.33\textwidth}
      \centering
      \includegraphics[width=1.\textwidth]{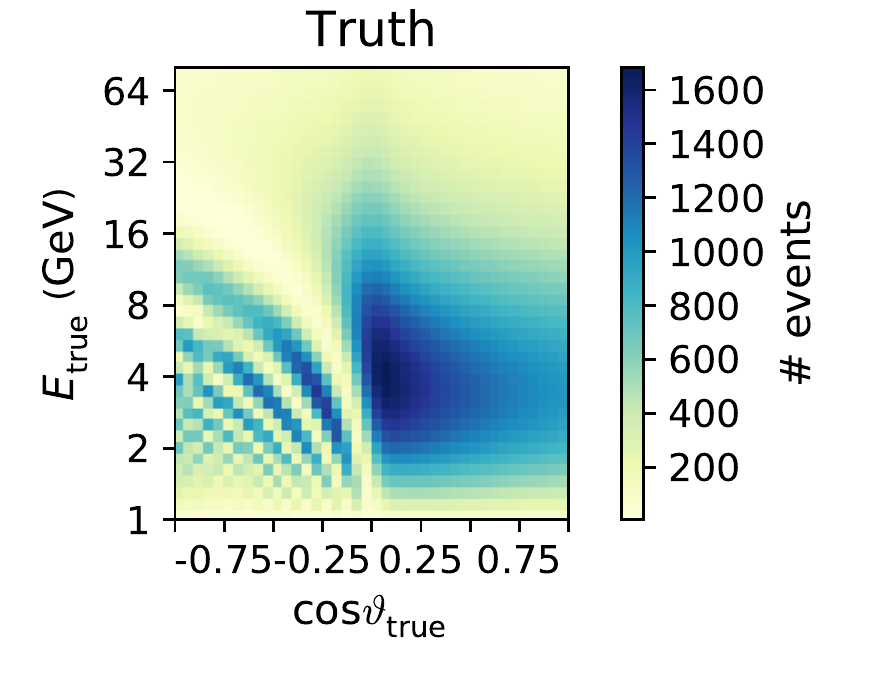}
  \end{minipage}%
  \begin{minipage}{.67\textwidth}
      \centering
      \includegraphics[width=1.\textwidth]{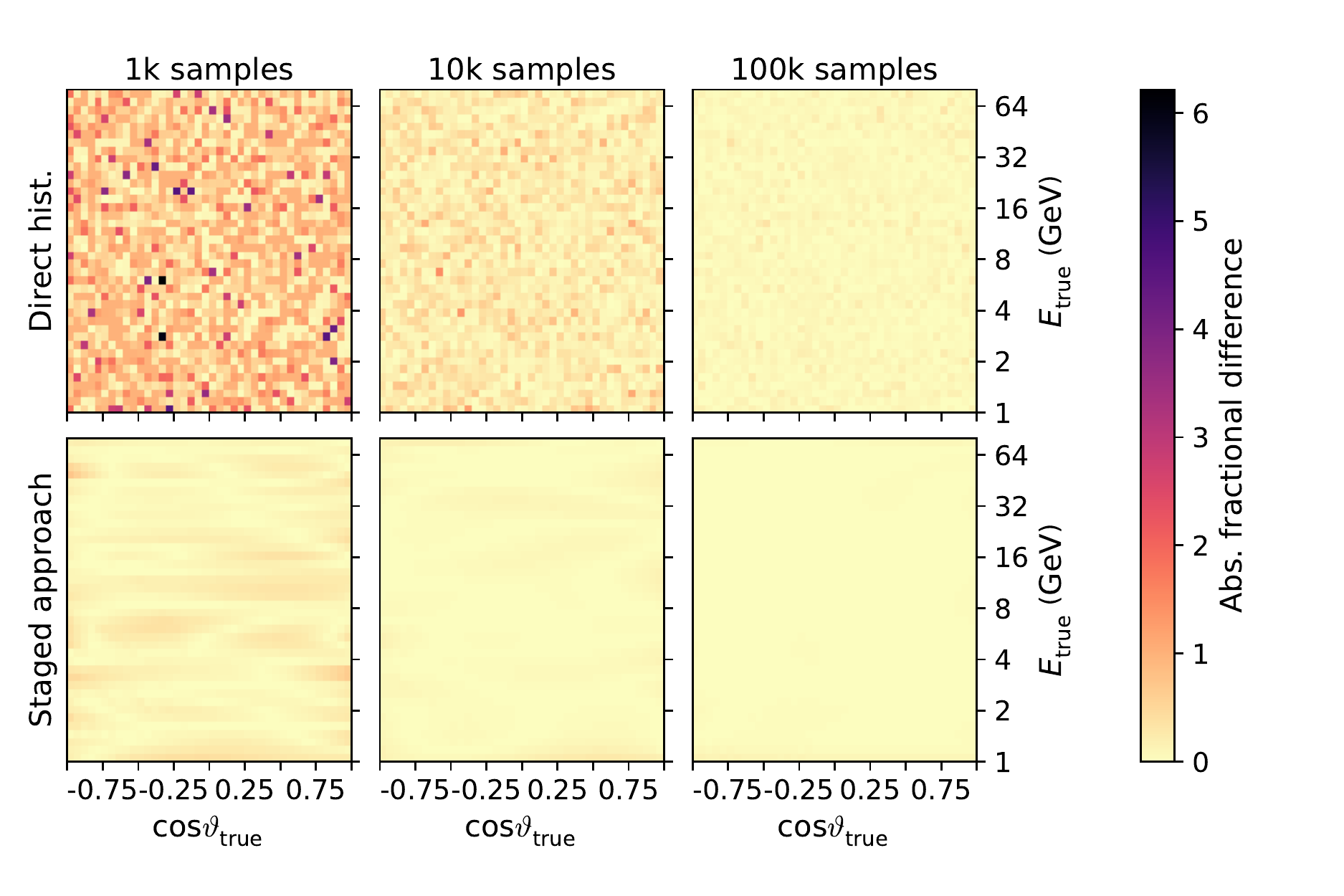}
  \end{minipage}%
  \caption{
  Parametric reference distribution after the first three stages (flux, oscillation, and detection) for the {\numucc} channel in ({\CZtrue}, {\Etrue}) (left panel) and relative residuals ($|N-N_{\rm true}|/N_{\rm true}$) for the direct histogramming (right panel, top row) and our proposed method (right panel, bottom row) on a $40\times40$ grid using different amounts of simulated events. (Note that the numbers are not percentages.)
  The three columns in the right panel represent different MC sample sizes of $10^3$, $10^4$, and $10^5$ events, respectively. The samples are drawn from the unbinned toy model distributions of \ref{sec:toy}.}
  \label{fig:aeff_vs_stats}
\end{figure}

Figure~\ref{fig:aeff_vs_stats} shows the truth template of {\numucc} events on a grid with $n_{\rm bins}=40 \times 40$ points together with the fractional deviations that arise when the same template is obtained from MC samples\footnote{Generated from the toy model in \ref{sec:toy}.} of different sizes using direct histogramming versus the smoothing method described above.
We use {\numucc} events as an example here and below.
Table~\ref{table:aeff} gives the average (binwise, i.e. per degree of freedom) $\chi^2$ values defined as
\begin{equation}
    \mean{\chi^2} = \frac{1}{n_{\rm bins}}\chi^2 = \frac{1}{n_{\rm bins}}\sum_{i=1}^{n_{\rm bins}} \frac{\left(\mu^\prime_i-\mu^{\mathrm{ref}}_i\right)^2}{\mu^{\mathrm{ref}}_i}
    \label{eq:average chi2}
\end{equation}
and maximal $\chi^2$ values defined as
\begin{equation}
    \chi^2_{\rm max} = \max_{1\,\leq\, i\,\leq\, n_{\rm bins}}\left[\frac{\left(\mu^\prime_i-\mu^{\mathrm{ref}}_i\right)^2}{\mu^{\mathrm{ref}}_i}\right]
    \label{eq:max chi2}
\end{equation}
by which the templates from the our method and from direct histogramming deviate from truth (with bin counts $\mu^\mathrm{ref}_i$).

The $\chi^2$ values provide direct insight into how the accuracy of the template description compares to the statistical uncertainty of the real data that would be observed. Since the observed data underlies Poisson fluctuations it has an average deviation from truth of $\chi^2=1$ per bin. An analysis of real data, however, can only test templates based on MC. These exhibit their own statistical uncertainties, resulting in finite $\chi^2$ deviations from truth, shown in Table~\ref{table:aeff} as a function of MC sample size. It is essential that these inaccuracies inherent to the template generation process are considerably smaller than the statistical fluctuations in data in order to ensure accurate statistical inference.

Applying our method we find deviations that are lower by a factor of about 40 for the smallest MC set, and by a factor of about 13 for the largest.
It is noteworthy that the maximum deviation ($\chi^2_{\rm max}$) across all bins decreases monotonically with MC sample size, confirming that the used smoothing method does not introduce any observable bias.

\begin{table}[tb]
\centering
\begin{tabular}{lc|cccc}
\multicolumn{2}{l|}{Sample size}                   & $10^3$        & $10^4$     & $10^5$    & $10^6$ \\ \hline 
\multirow{2}{*}{Direct hist.} & $\mean{\chi^2}$     & 215     & 22.5   & 2.07   & 0.201  \\
                              & $\chi^2_{\rm max}$      & 21600   & 1810 & 79.4  & 11.2 \\[6pt]
\multirow{2}{*}{Staged approach} & $\mean{\chi^2}$  & 5.14       & 0.526    & 0.0615   & 0.0156 \\
                              & $\chi^2_{\rm max}$      & 460     & 17.2   & 2.27   & 0.975   \\
\end{tabular}
\caption{Average $\chi^2$ per bin and the worst-case bin's $\chi^2$ value comparing templates on a $40\times40$ grid in (\Etrue, \CZtrue)-space (i.e., before applying reconstruction resolutions) generated by direct histogramming (top) and the smoothed-staged approach (bottom) with the toy model's reference template.
Shown are values obtained for independent input MC samples of various sizes (from $10^3$ up to $10^6$ events per flavor/interaction type).}
\label{table:aeff}
\end{table}

\subsection{Reconstruction}
\label{sec:reco}

The usual way to obtain templates in the space of reconstructed variables is to place each individual MC event in the final-level distributions according to the reconstruction information that the event carries. This is the case for both methods that are used for comparison: direct histogramming and direct KDE, the only difference between these being how the final-level distributions are estimated. While this approach correctly takes into account joint dependencies of the event reconstruction on the involved variables, it is particularly sensitive to small MC sample sizes due to the potentially high dimensionality of the space of reconstructed variables.
In contrast, the staged approach uses the available MC simulation to construct detector resolution functions which we integrate to form a transformation that maps a template in true variables (such as that shown on the left in Figure~\ref{fig:aeff_vs_stats}) onto the space of reconstructed variables, what we refer to as the final-level template.

In the case study of the NMO analysis, the {mapping} of true variables ({\Etrue} and {\CZtrue}) to reconstructed variables ({\Ereco}, {\CZreco}, and event class) is extracted from the MC as a ``migration'' tensor of order five, $\mathcal{M}_{ijklm}$. It maps the histogram of event counts in the two-dimensional space of true variables, $h_{ij}$, to the observed histogram of event counts in the three-dimensional space of reconstructed variables, $h^\prime_{klm}$:
\begin{equation}
h^\prime_{klm} = \sum_{i,j}\mathcal{M}_{ijklm} h_{ij}\text{ .}
\end{equation}

The reconstruction transform in general has to be computed as a five-dimensional transform, as all five dimensions can depend on one another---i.e. they are correlated.
Studying the correlations among the dimensions in our particular MC revealed, however, that $\Ereco$ only depends on event class and $\Etrue$, $\CZreco$ depends on event class and both input dimensions, and event class only depends on $\Etrue$.
For each of the three reconstruction variables, we subdivide the MC in the quantity's dependent dimensions to the point that correlations are not visible and that all events in the subdivision can be assumed to be samples from the same one-dimensional distribution.---i.e. the resolution functions we generate.


There is a trade-off in terms of how much to subdivide the MC for producing these resolution functions.
Since resolution changes as a function of a dependent dimension, sufficiently narrow subdivisions in that dimension group together MC events drawn from essentially the same distribution.
Subdivisions that are too wide will group together events drawn from different distributions and the resulting resolution functions will be erroneous.
However, narrower subdivisions admit fewer MC events in each subdivision and so lead to greater statistical variations in the estimated resolution functions (i.e., their shapes will be more affected by random fluctuations in the MC).

To balance these competing factors, we devised the following heuristic.
For the quantity being characterized, we divide each dependent dimension evenly---except event class, which is binary.
$\Etrue$ is divided evenly in log-space to help ensure even subdivisions group together events with similar energy resolution, as this quantity changes more rapidly at low $\Etrue$ than at high $\Etrue$.
We allow each subdivision of $\Etrue$ to separately expand enough to capture at least 100 events, and at least 500 events in each subdivision of $\CZtrue$.
If expansion is performed, subdivisions' upper and lower edges are expanded by the same factor (up to the limits of the dimension).
The captured events are then used to produce resolution functions.

The remaining parameters that require tuning in this heuristic are the number of subdivisions to use for each dependent dimension for each quantity being characterized.
For this, we visually inspect the 2-dimensional distributions of each characterized quantity as a function of each dependent dimension and require that the events in each subdivision do not display strong dependence on the dependent dimension.

\begin{figure}[tb]
  \begin{center}
    \includegraphics[height=2in]{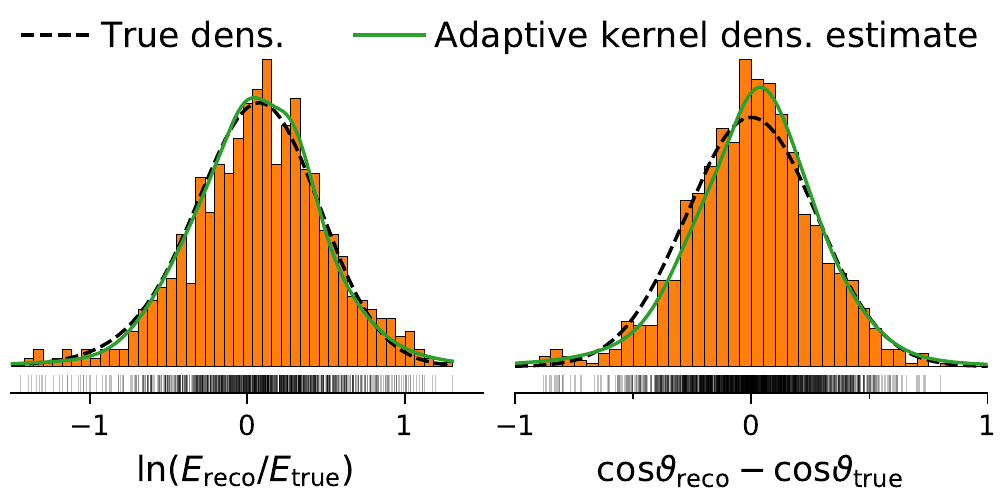}
  \end{center}
  \caption{
  Example energy and cosine-zenith-angle resolution distributions for {\numucc} events classified as cascades, estimated with histograms and adaptive KDE.
  Energy resolution is shown for 100 events with $\Etrue \in [26.7, 29.8]\,{\rm GeV}$ and cosine-zenith resolution for 100 events with $\Etrue \in [1.0, 1.1]\,{\rm GeV}$.
  The samples used to construct the histogram and KDE are shown by vertical lines beneath the histograms.
  }
  \label{fig:vbwkde_example}
\end{figure}

If the functional form of the resolution functions is known, a parametric model of this form fit to the MC yields the most accurate and lowest variance reconstruction transform. 
However, as we do not know the form of these functions, a non-parametric density estimation technique is used to approximate them.
In particular, we chose to use adaptive KDE~\cite{abramson1982} with bandwidths scaled uniformly such that the narrowest is that found from the Improved Sheather Jones (ISJ) algorithm~\cite{botev2010}.
KDE works by placing a kernel function (we use a Gaussian) centered at the value of each event's variable to be described and then summing over all kernels.
Adaptive bandwidth KDE uses different widths for each kernel, where the bandwidths are inversely proportional to the density of points near the location of the kernel.
The ISJ bandwidth selection algorithm used to normalize the kernel widths is an improvement over predecessor algorithms (e.g.,~\cite{10.2307/2345597,scott}) in that it does not make assumptions that the quantity being estimated is drawn from a Gaussian distribution.
In our experience, this outperforms fixed bandwidth KDE by not underestimating the heavy-tailed distributions we encounter, but it bears repeating that other density estimation techniques can yield better or worse results depending on the specifics of the MC in question.
An example of two resolution functions (one for both energy and zenith angle, respectively) estimated using the adaptive KDE method is shown in Figure~\ref{fig:vbwkde_example}.

\begin{figure}[tb]
  \begin{minipage}{.33\textwidth}
      \centering
      \includegraphics[width=1.\textwidth]{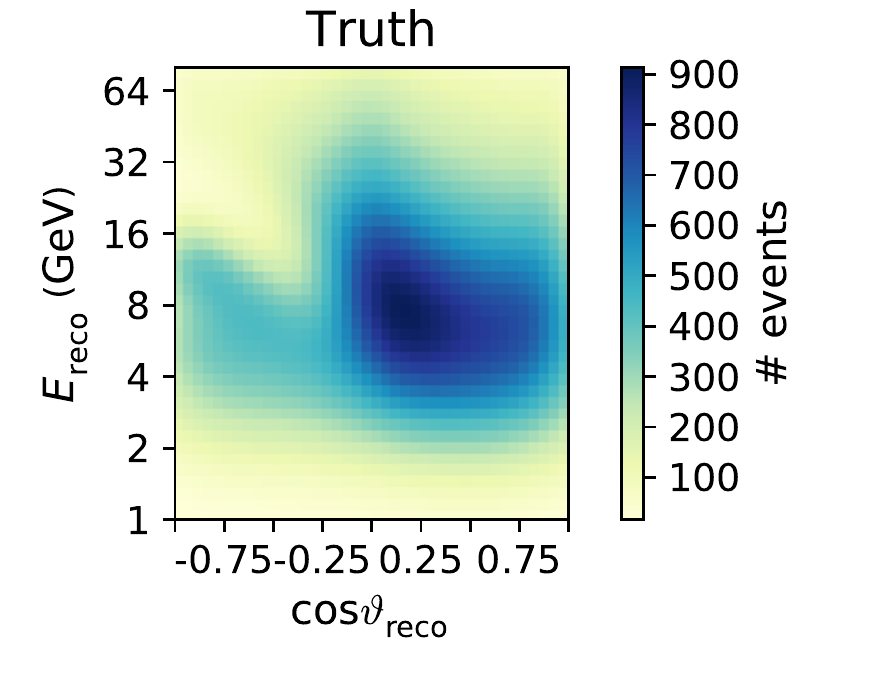}
  \end{minipage}%
  \begin{minipage}{.67\textwidth}
      \centering
      \includegraphics[width=1.0\textwidth]{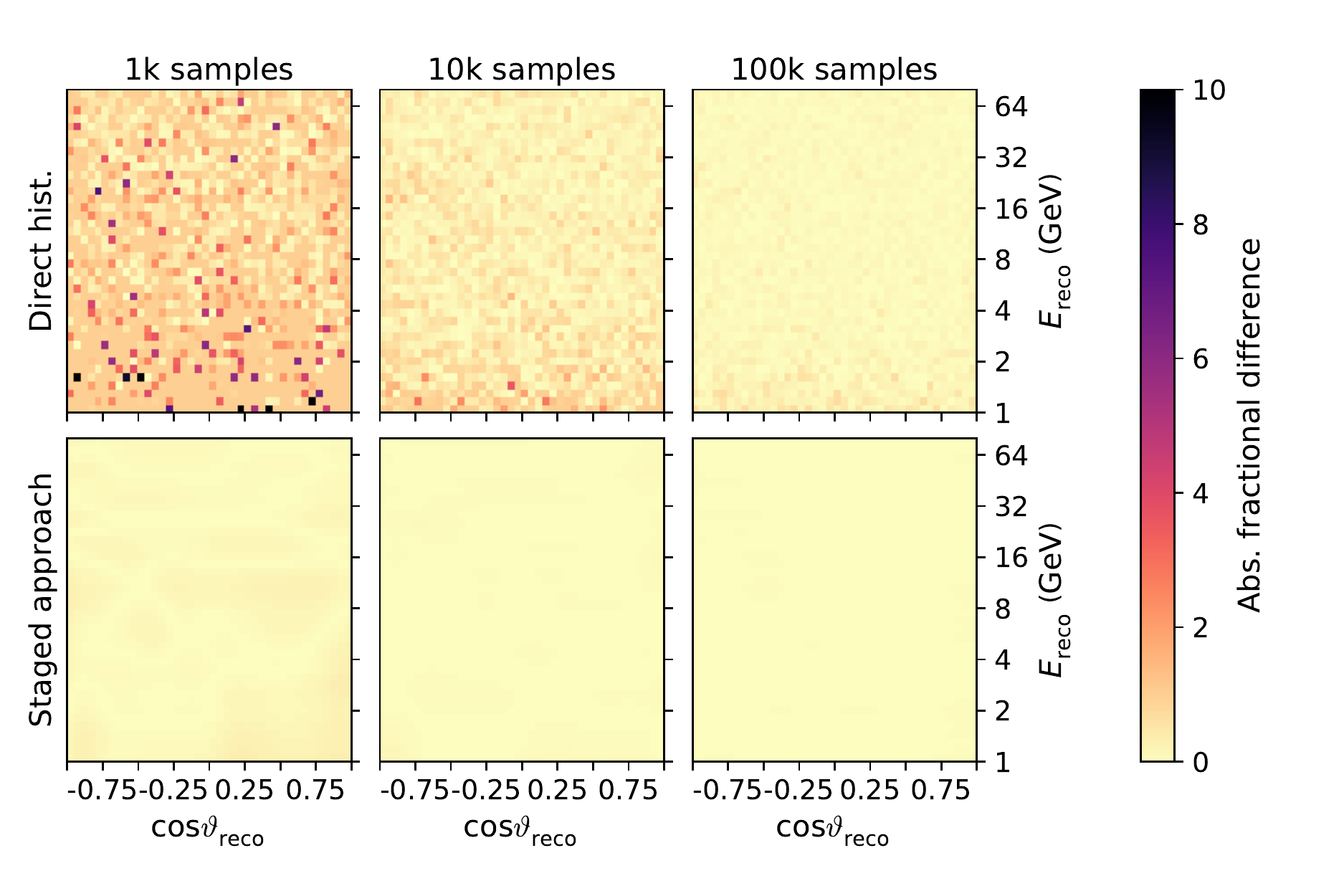}
  \end{minipage}%
  \caption{
  Same as Figure~\ref{fig:aeff_vs_stats}, but comparing final-level templates after all four stages are applied.
  Note that the residuals in the 1k-samples plot for direct histogramming go up to 31 but the scale is clipped at 10.
  }
  \label{fig:reco_vs_stats}
\end{figure}

Figure~\ref{fig:reco_vs_stats} again demonstrates that templates obtained from our KDE-based reconstruction stage deviate much less from the parametric reference template after reconstruction than templates from direct histogramming of reconstructed MC events.

\section{Validation and Comparison of Templates}
\label{sec:valid}
This section more closely examines the templates generated with the staged approach and compares them---along with those generated by the other two methods (histograms and KDE)---to the parametric reference distributions of the toy detector model. This validation is split into two parts. The first examines the grid of points that are used to numerically approximate the integral over the first three stages, whereas the effect of smoothing is investigated in the second.

\subsection{Sampling Grid}
\label{subsec:sampling grid}
In order to demonstrate the validity of our choice of grid points shown in Table~\ref{table:stage binning} as well as the equivalence between the staged approach and traditional MC weighting as grid point spacing in {\Etrue} and {\CZtrue} is reduced, we compare the staged approach without smoothing (i.e. using raw histograms as transforms in place of smoothing functions and KDEs) to direct histogramming.
The specific comparison done here without smoothing is solely for the purpose of validating the principle of stages vs. direct histograms.

Table~\ref{table:hist_valid} shows the $\chi^2$ difference (cf. Equations~(\ref{eq:average chi2})~and~(\ref{eq:max chi2})) 
%
between the final templates obtained from the staged approach (with bin counts $\mu^\prime_i$) and direct histogramming ($\mu^\mathrm{ref}_i$) for a variety of grid point densities in {\Etrue} and {\CZtrue}, using the same MC set of size $10^6$ for both methods.
These templates are output with a binning of $40 \times 40 \times 2$ in {\Ereco}, {\CZreco}, and event class.
The relative decrease in the average $\chi^2$ value roughly scales with the inverse of the relative grid density increase, thus confirming that the two methods will agree to arbitrary precision in the asymptotic limit.
In the following, for practical reasons we limit ourselves to the specific case summarized in Table~\ref{table:stage binning}.

\begin{table}[tb]
\centering
\begin{tabular}{c|cccccc}
                            Grid ($M\times N$)   & $40\times 40$   & $80\times 80$   &$160\times 160$ & $320\times 320$ & $640\times 640$ & $1280\times 1280$ \\ \hline
$\mean{\chi^2}$                    & 0.01067   & 0.00253   & 0.00060   & 0.00014   & 0.00003   & 0.00001 \\
$\chi^2_{\rm max}$                  & 1.45906   & 0.46930   & 0.19718   & 0.04974   & 0.00634   & 0.00172 \\[6pt]
\end{tabular}
\caption{
    Average and maximal $\chi^2$ deviations per bin of the final $40\times40\times2$ binning between final templates of non-smoothed staged approach and direct histogramming, for different grid point densities in ({\Etrue}, {\CZtrue}) for the first three stages, using an MC sample of $10^6$ events. The last (=reconstruction) stage uses a reduced binning, as described in the text.}
    
\label{table:hist_valid}
\end{table}

\subsection{Smoothing}
\label{subsec:smoothing}
To validate the final templates with smoothing applied at each stage, we compare them directly to truth.
For reference, we also show the agreement resulting from both the direct histogramming and the direct KDE methods.

\begin{figure}[tb]
  \begin{minipage}{.5\textwidth}
    \centering
    \includegraphics[width=1.0\textwidth]{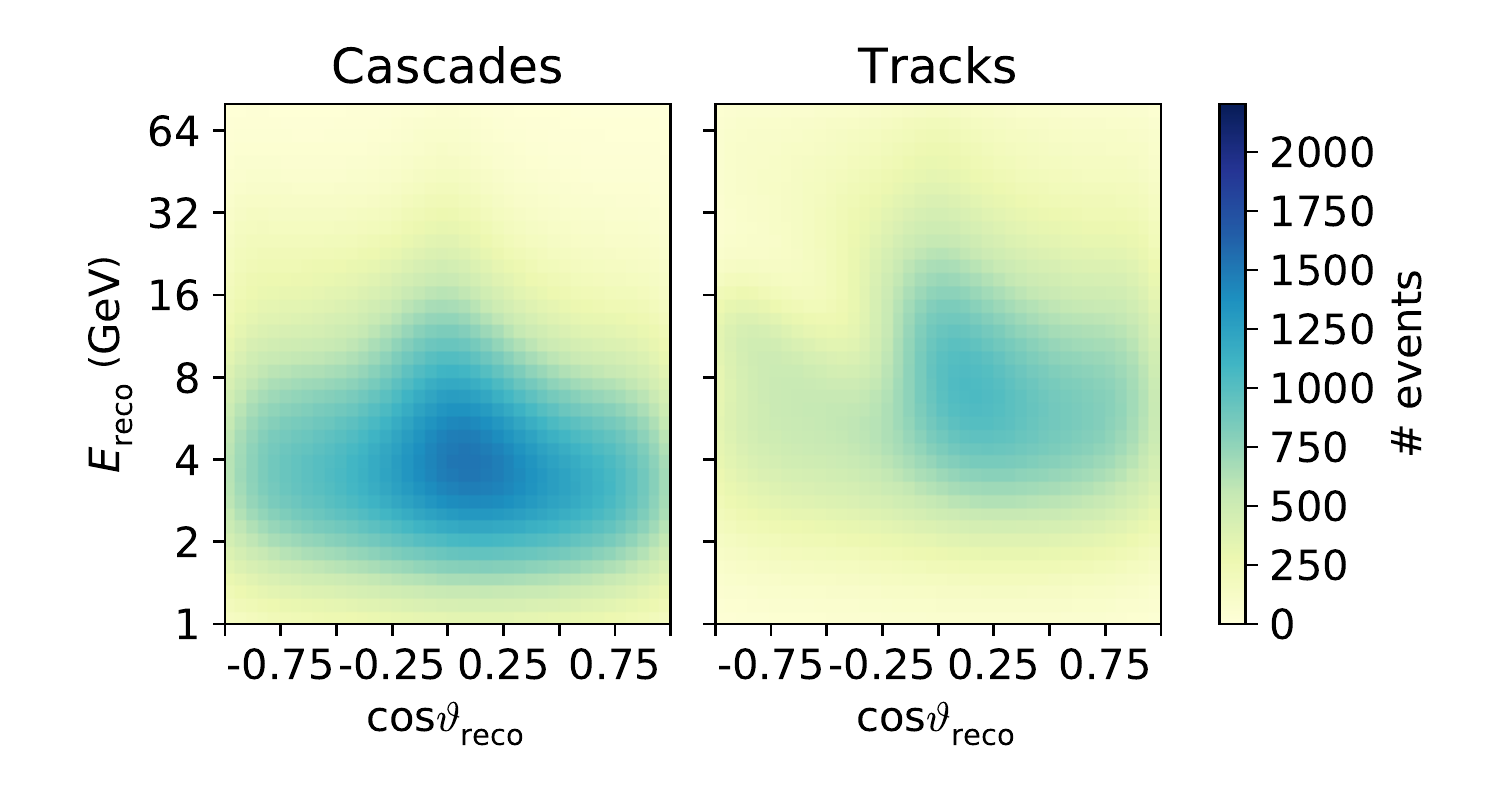}
    \vspace{-20pt}
    \subcaption{Truth}
    \vspace{6pt}
  \end{minipage}
  \begin{minipage}{.5\textwidth}
    \centering
    \includegraphics[width=1.0\textwidth]{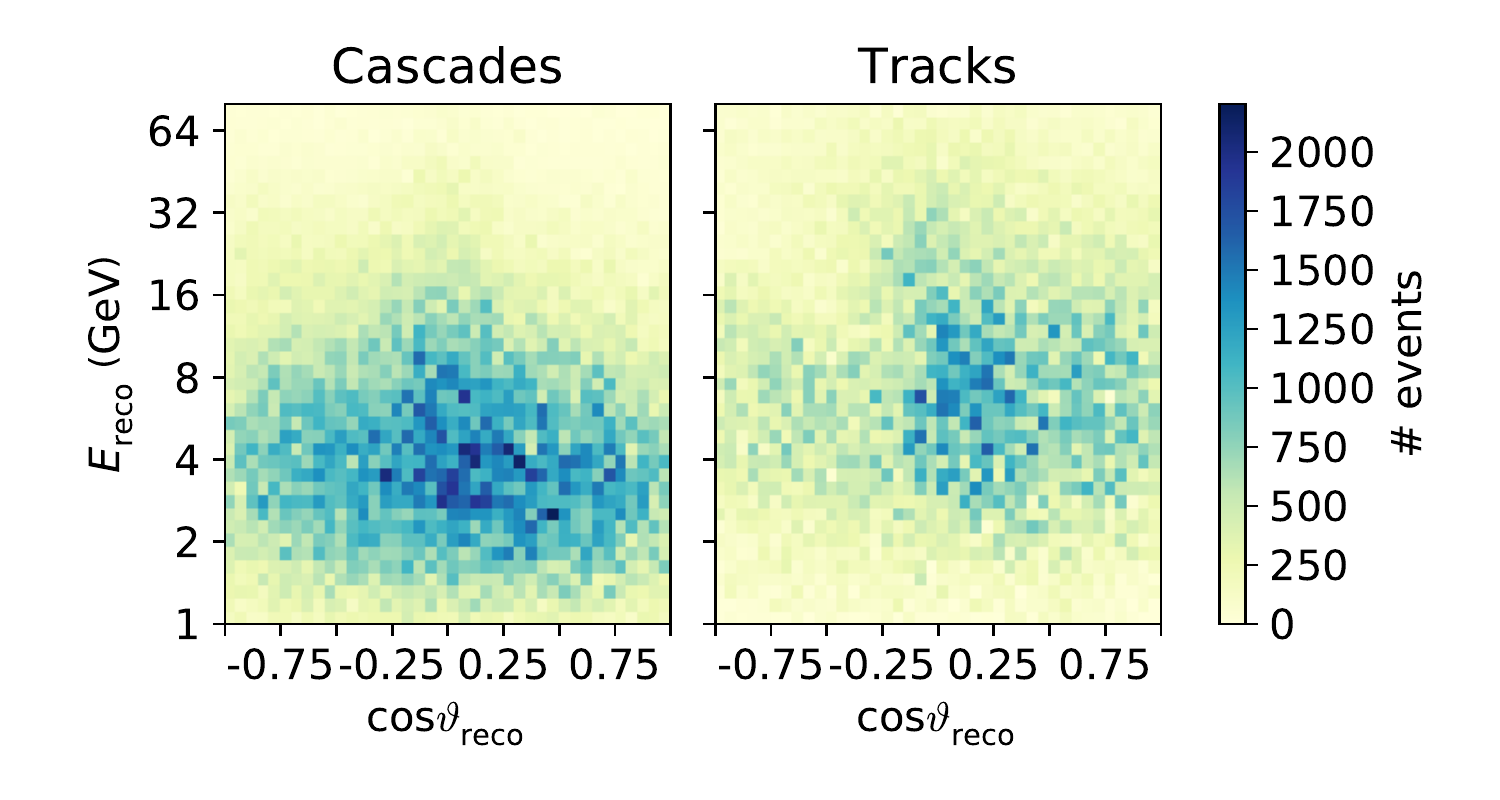}
    \vspace{-20pt}
    \subcaption{Direct Histogramming}
    \vspace{6pt}
  \end{minipage}
  \begin{minipage}{.5\textwidth}
    \centering
    \includegraphics[width=1.0\textwidth]{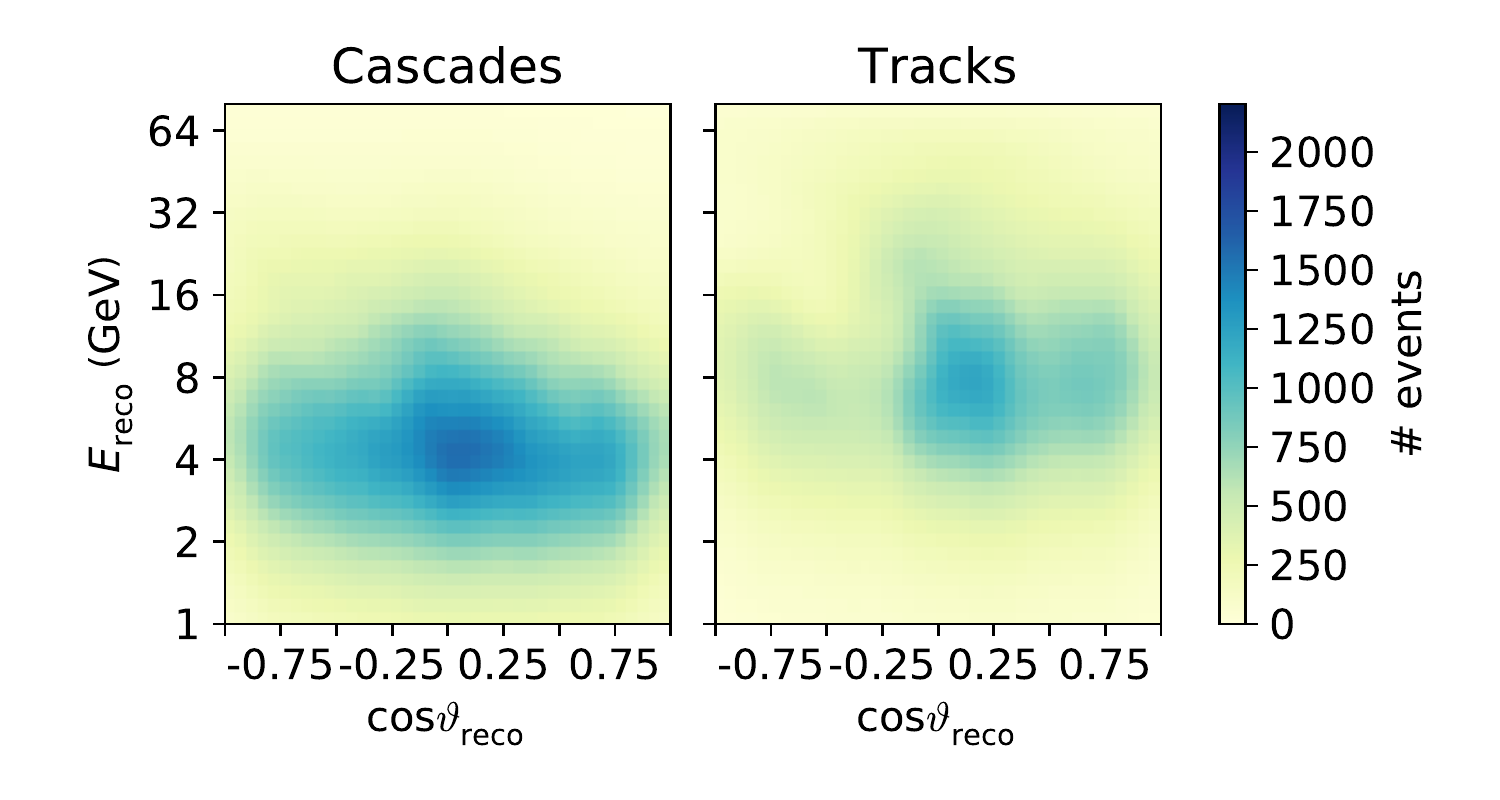}
    \vspace{-20pt}
    \subcaption{Direct KDE}
  \end{minipage}
  \begin{minipage}{.5\textwidth}
    \centering
    \includegraphics[width=1.0\textwidth]{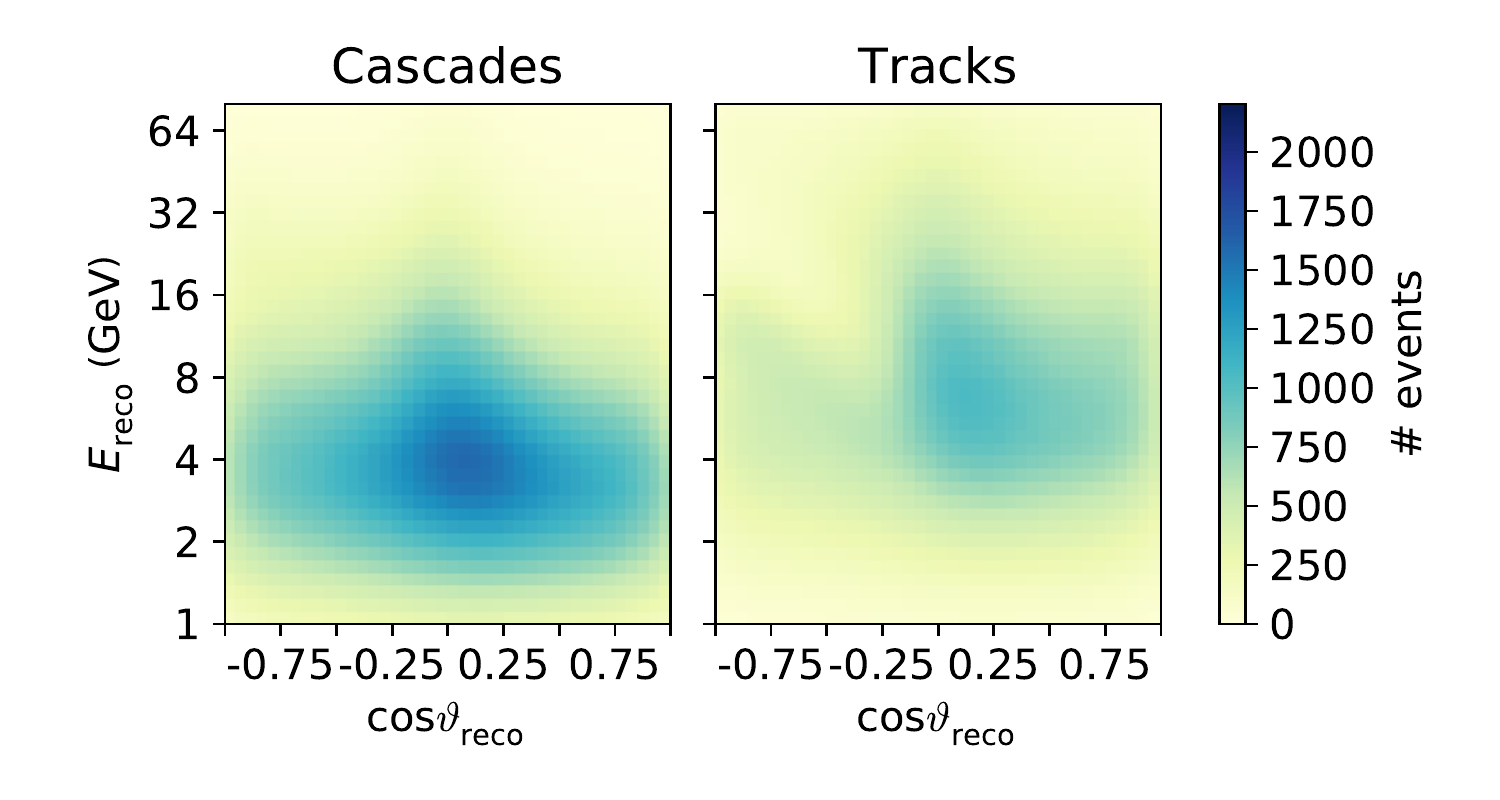}
    \vspace{-20pt}
    \subcaption{Staged Approach}
  \end{minipage}
  \caption{Final-level templates used for the example data analysis. The reference distributions (truth) obtained directly for the toy detector model parameterizations are shown in panel (a). Given the same sample of $10^4$ events the estimated distributions using histograms are shown in panel (b), using KDEs in panel (c), and using the staged approach in panel (d).}
\label{fig:valid}
\end{figure}

While Table~\ref{table:valid} quantifies deviations from the reference distributions again in terms of $\chi^2$ and in dependence of MC sample size, Figure~\ref{fig:valid} displays the final-level templates for each of the aforementioned methods using a sample with $10^4$ events.

The staged approach outperforms the two alternatives in terms of $\chi^2$ values by more than one order of magnitude for all those sample sizes studied here.
Furthermore, inaccuracies of the templates from the staged approach scale with the inverse of sample size almost as fast as those of templates from direct histogramming.
In addition, it is noteworthy that the KDE method shows comparably slow convergence, i.e., it performs worse than direct histogramming for the sample size of $10^6$.


While for the experimental data (or pseudo-data) one expects statistical fluctuations on the order of $\chi^2 = 1.0$ per bin, the accuracy of the templates must be better than this.
As shown in Table~\ref{table:valid}, considering a sample size of $10^4$ and the staged approach, the average $\chi^2$ deviation from truth (using the same $\chi^2$ definition as for data) is only about 30\% of what is expected just from statistical fluctuations in data, while more than $10^6$ events would be necessary to achieve the same average $\chi^2$ using direct histogramming or KDE.
(See Table~\ref{table:valid} for details.)
Therefore, to reach an equal accuracy, two or more orders of magnitude larger samples are needed for histogramming or KDE compared to the staged approach.
The next section illustrates the implications for running a data analysis.

\begin{table}[h]
\centering
\begin{tabular}{lc|cccc}
 \multicolumn{2}{l|}{Sample size}                   & $10^3$        & $10^4$    & $10^5$    & $10^6$ \\ \hline 
\multirow{2}{*}{Direct Histogramming} & $\mean{\chi^2}$     & 468     & 42.6  & 4.27   & 0.458  \\
                              & $\chi^2_{\rm max}$      & $3.4\cdot10^4$  & 906 & 138 & 10.5 \\[6pt]
\multirow{2}{*}{Direct KDE}   & $\mean{\chi^2}$     & 32.2      & 11.4  & 3.67   & 1.25   \\
                              & $\chi^2_{\rm max}$      & 245     & 90.2  & 50.3  & 25.3 \\[6pt]
\multirow{2}{*}{Staged Approach} & $\mean{\chi^2}$  & 3.01       & 0.303   & 0.111   & 0.0301   \\
                              & $\chi^2_{\rm max}$      & 47.4      & 3.03   & 1.80   & 0.387   \\

\end{tabular}
\caption{
    Average and maximal $\chi^2$ deviations per bin of the final $40\times40\times2$ binning between final templates of the three shown methods and truth, for independent input MC samples of various sizes. Note that the staged approach has smoothing applied (the default), in contrast to Table~\ref{table:hist_valid}.}
\label{table:valid}
\end{table}

\section{Example Analysis Results}
\label{sec:res}

\begin{figure}[th]
  \centering
  \includegraphics[height=3.0in]{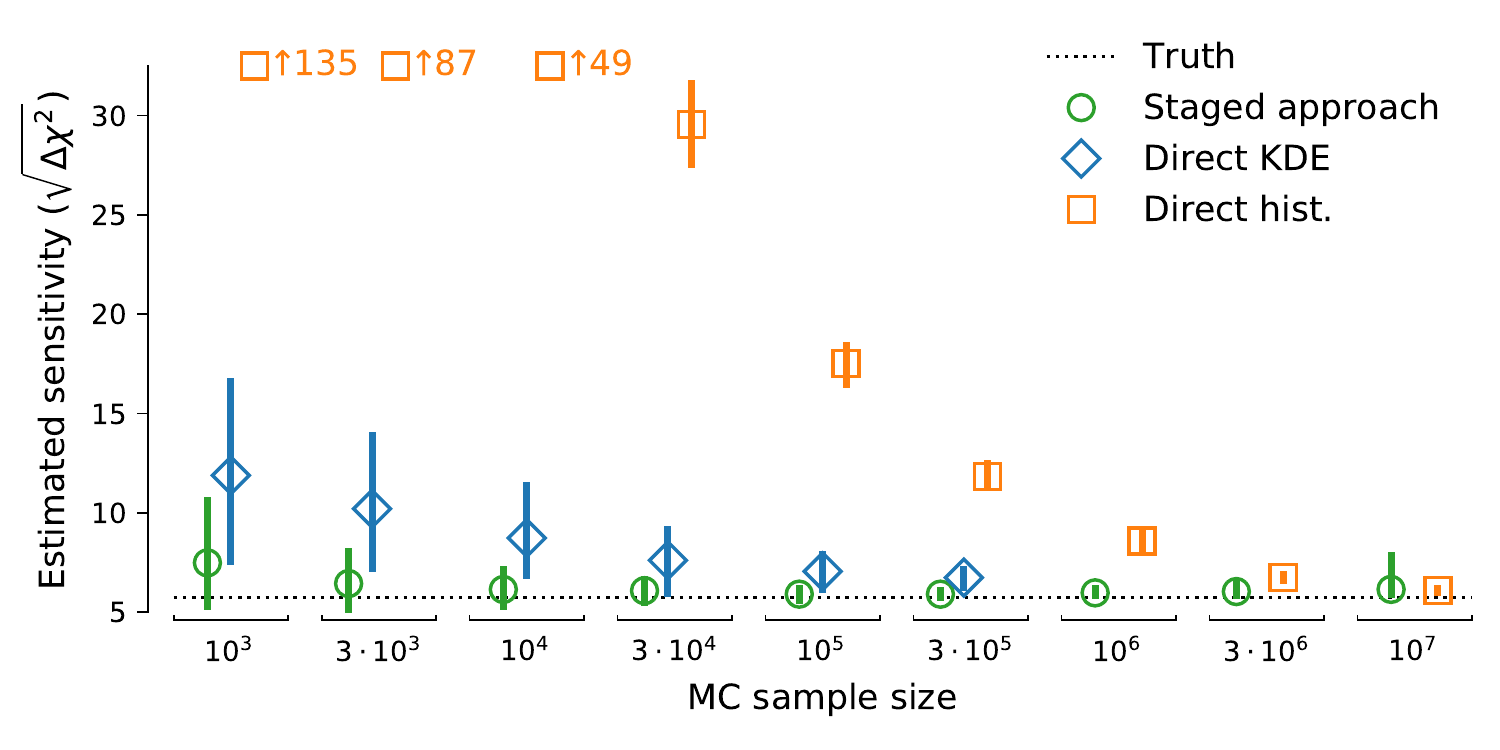}
  \caption{
  Estimated sensitivity ($\sqrt{\Delta\chi^2}$) to the NMO vs.~sample size for direct histogramming, direct KDE, and the proposed staged smoothing methods applied to multiple (between 50 and 200) statistically independent toy MC sets.
  Vertical lines indicate central 68\% quantiles of the ensemble.
  The dashed horizontal line shows the significance obtained from truth templates based on the parametric toy detector model.
  The staged approach outperforms the other methods---both in terms of bias and variance---for sample sizes through $3\cdot10^6$, with direct histogramming only matching the staged approach using $10^7$ samples.
  Note that no data points exist for direct KDE and sample sizes above $3\cdot10^5$, as computational processing times become impractically large.
  Also note that direct histogramming is off-scale high for fewer than $3{\cdot}10^4$ events (mean values are indicated to the right of the corresponding markers).
  }
  \label{fig:nmo_result}
\end{figure}

To illustrate the impact of sample size, we show the resulting $\sqrt{\Delta\chi^2}$ as an estimate for the sensitivity to the NMO for our example analysis in Figure~\ref{fig:nmo_result}.
For reference, the true result is derived directly from the exact templates based on the parametric toy detector model and lies at $\sqrt{\Delta\chi^2}=5.75$.
For the three methods discussed throughout this paper, the statistical uncertainty of the obtained sensitivity is indicated by error bars in the figure.
This uncertainty is computed from several statistically independent MC sets\footnote{Each MC set is used together with the staged approach to generate one Asimov toy data template and $\mathcal{O}(10^3)$ ``test" templates.} and indicates the central 68\% quantile of each ensemble. In particular, as the sensitivity proxy does not take into account MC uncertainty~\cite{Glusenkamp:2019uir,Arguelles:2019izp}, this range is not, a priori, expected to reflect any sensitivity bias for the three methods. 

The uncertainty reveals that the methods exhibit quite different intrinsic fluctuation of their respective sensitivity estimates, as well as different scaling behavior of the variance with sample size.
As sample size decreases, direct histogramming without any smoothing applied results in an increasing overestimation of a {\vlvnt}'s ability to exclude the wrong neutrino mass ordering.
In the most extreme case shown here (corresponding to the smallest sample size of $10^3$
), the sensitivity is estimated to be more than one order of magnitude greater than the actual capability of the experiment.
Only for the sample size of $10^7$ does direct histogramming indeed give reliable results. This is expected from the simple rule of thumb (cf. Section~\ref{subsec: MC gen requirements}) of $\mathcal{O}(10^{4})$ events per bin $\times\ \mathcal{O}(10^{3})$ bins. 

Illustratively, an undersampling of the detector response distributions due to low MC statistics is highly likely to lead to an overestimation of the experiment's sensitivity because the NMO signature that is present in the space of true variables is carried over to random bins in the reconstructed observables with reduced cancellation\footnote{For example, if a bin in the final-level template is solely populated by (unweighted) MC neutrinos, and no anti-neutrinos, or vice-versa, it will contribute artificially strong to the overall NMO sensitivity due to the missing summation over both event types (cf. Section~\ref{sec:nmo}).}.

Applying KDE smoothing to the weighted events instead of histogramming them (i.e., direct KDE) leads to a reduction of the overestimated sensitivity for sample sizes of up to at least $3\cdot 10^5$ but is not able to eliminate the bias entirely for the tested sample sizes.
For sample sizes larger than $\mathcal{O}(10^{5})$, the direct KDE method is too computationally expensive to deliver results within a reasonable time (for more details on timing, see Section~\ref{sec:bench}).

The estimated sensitivity using the staged approach is statistically compatible with the true sensitivity across the whole range of sample sizes considered.
It shows no bias and lower variance for predicting sensitivity to physics compared to the other methods within the limits of our testing.

\section{Benchmarks}
\label{sec:bench}

\begin{figure}[ht]
  \begin{center}
    \includegraphics[width=0.7\textwidth]{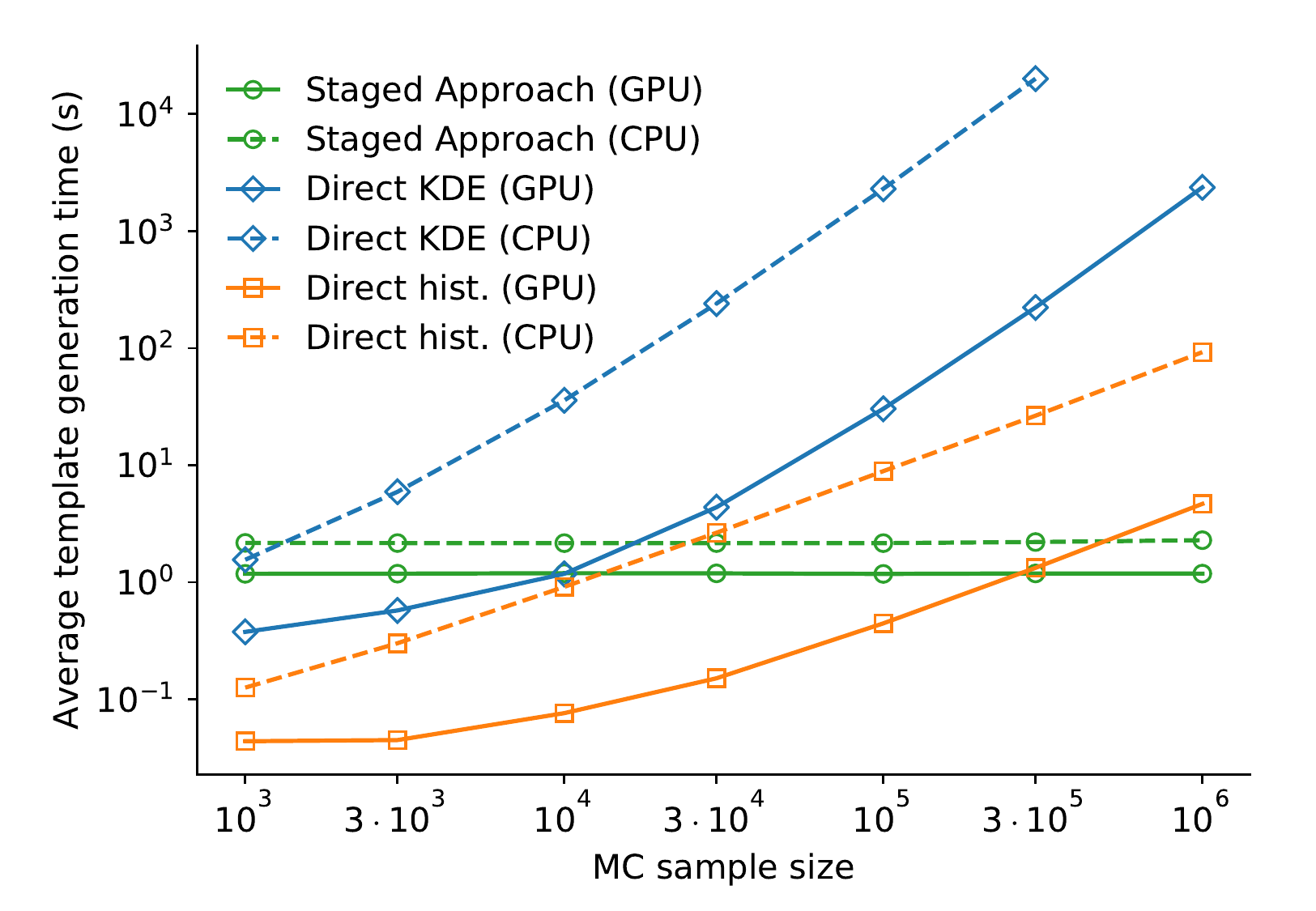}
  \end{center}
  \vspace{-15pt}
  \caption{Average template generation time during a typical analysis for input datasets of varying size, shown for the direct histogramming, the direct KDE, and the staged approach. Solid lines represent timings based on (partial) GPU acceleration, whereas the dashed ones are for CPU-only calculations.}
  \label{fig:timings}
\end{figure}

Whether a given analysis method is useful in a realistic setting depends not only on its numerical reliability, but also on how long it takes to compute the quantity of interest (note that this duration is in addition to the initial time needed to generate the MC).
For reference, we performed benchmarks of the template generation times in the course of a typical analysis process\footnote{Timings were obtained on a computer with an Intel Xeon E5-1660 v3 3.0 GHz CPU and an NVIDIA GeForce GTX Titan X GPU.}.
These are compiled in Figure~\ref{fig:timings}.

Note that no initial start-up times---such as the construction of the smearing kernels used within the reconstruction stage---are included here.
For all three methods separate timings based on our CPU-only and GPU-accelerated implementations are provided.

While for sample sizes below $10^4$ to $10^5$ events direct histogramming is the fastest method, it is unusable owing to the large fluctuations associated with the templates it produces, which in turn result in the grossly overestimated sensitivites shown in Figure~\ref{fig:nmo_result}.
Direct KDE only proves competitive when used in conjunction with the smallest datasets.
The faster-than-linear scaling of its computational needs with sample size then quickly renders it impractical to use.
Our proposed method is independent of sample size by construction (excluding initial start-up costs), but will get more expensive if a finer grid point spacing is desired.

The timing difference between the CPU and GPU implementation of the staged approach is not as large as for the other methods, since it is only using the GPU for parallelization of the neutrino oscillation weights calculation (whereas the other methods make use of the GPU more extensively).

\section{Summary}
\label{sec:summary}

The search for small physics effects in high statistics neutrino oscillation experiments requires careful treatment and use of simulated data.
Statistical fluctuations within distributions obtained from Monte Carlo simulations are able to severely distort an analysis, rendering derived constraints or sensitivities essentially meaningless.

The staged approach we have presented serves two main purposes. Firstly, computational expense is reduced through sampling of physics and detector response distributions on a discrete grid instead of computing a weight for every individual Monte Carlo event.
In this respect, we have demonstrated that our method of breaking down the template generation into independent stages converges to the MC weighting scheme when using a grid of a high enough, albeit feasible, density.
For a fixed number of grid points, the template generation time has been shown to be independent of the input sample size.
Secondly, the staged approach allows the application of smoothing techniques to a detector's response functions.
In the specific example shown, the detection stage characterizes the detector's effective area by integrating weighted MC events on a grid and smoothing the resulting histogram, followed by the event reconstruction stage using an adaptive KDE smoothing on the resolution functions applied to arrive at final-level templates.
This has proven superior to the smoothing of the final event distributions since it is faster and---even more importantly---yields more accurate and robust results.
The presented choice of smoothing techniques works sufficiently well for our purposes, but this choice is neither unique nor do we claim it to be optimal, and it depends on the wider experimental context. Beside this choice, our overall approach may prove particularly useful when a fast assessment of the physics potential of various detector designs is desired, or when analysis methodologies are optimized. Any final-level analysis will likely rely on large quantities of MC to guarantee the precise and accurate modelling of the experiment.

In the example neutrino mass ordering analysis that we have conducted---to benchmark and compare the different approaches---we found that direct histogramming of events leads to a gross overestimation of sensitivities when used in conjunction with small numbers of events ($\lesssim 10^6$ events for our toy model).
Conversely, the proposed staged approach leads to correct results that are largely unaffected by the sample size across the tested range and the variance of results is small compared to the result above about $10^4$ neutrino events.
This means that the necessary amount of simulated events is reduced significantly (by about two orders of magnitude in our example)---an important aspect especially since Monte Carlo event simulation and reconstruction times can represent major hurdles to progress in the field of neutrino oscillation experiments.

\section*{Acknowledgments}
The IceCube collaboration gratefully acknowledges the support from the following agencies and institutions:
USA {\textendash} U.S. National Science Foundation-Office of Polar Programs,
U.S. National Science Foundation-Physics Division,
Wisconsin Alumni Research Foundation,
Center for High Throughput Computing (CHTC) at the University of Wisconsin-Madison,
Open Science Grid (OSG),
Extreme Science and Engineering Discovery Environment (XSEDE),
U.S. Department of Energy-National Energy Research Scientific Computing Center,
Particle astrophysics research computing center at the University of Maryland,
Institute for Cyber-Enabled Research at Michigan State University,
and Astroparticle physics computational facility at Marquette University;
Belgium {\textendash} Funds for Scientific Research (FRS-FNRS and FWO),
FWO Odysseus and Big Science programmes,
and Belgian Federal Science Policy Office (Belspo);
Germany {\textendash} Bundesministerium f{\"u}r Bildung und Forschung (BMBF),
Deutsche Forschungsgemeinschaft (DFG),
Helmholtz Alliance for Astroparticle Physics (HAP),
Initiative and Networking Fund of the Helmholtz Association,
Deutsches Elektronen Synchrotron (DESY),
and High Performance Computing cluster of the RWTH Aachen;
Sweden {\textendash} Swedish Research Council,
Swedish Polar Research Secretariat,
Swedish National Infrastructure for Computing (SNIC),
and Knut and Alice Wallenberg Foundation;
Australia {\textendash} Australian Research Council;
Canada {\textendash} Natural Sciences and Engineering Research Council of Canada,
Calcul Qu{\'e}bec, Compute Ontario, Canada Foundation for Innovation, WestGrid, and Compute Canada;
Denmark {\textendash} Villum Fonden, Danish National Research Foundation (DNRF), Carlsberg Foundation;
New Zealand {\textendash} Marsden Fund;
Japan {\textendash} Japan Society for Promotion of Science (JSPS)
and Institute for Global Prominent Research (IGPR) of Chiba University;
Korea {\textendash} National Research Foundation of Korea (NRF);
Switzerland {\textendash} Swiss National Science Foundation (SNSF);
United Kingdom {\textendash} Department of Physics, University of Oxford.


\appendix

\section{Toy Data Model}
\label{sec:toy}
In the following we provide a parametric toy detector model used to transform the oscillated atmospheric fluxes into event counts. The functions we use either serve as direct inputs (truth) to the various stages of the simulation chain laid out in Section~\ref{sec:stages}, or as sampling distributions from which toy MC samples are drawn. We point out here that these are entirely empirically motivated, and should only be seen as proxies of the performance indicators in next-generation {\vlvnt}s (such as the IceCube Upgrade~\cite{upgrade_icrc:2019}, PINGU~\cite{TheIceCube-Gen2:2016cap,PINGU-LOI}, or KM3NeT/ORCA~\cite{Adrian-Martinez:2016fdl}).

Simplifications or limitations of the model do not affect the computational analysis techniques themselves. Rather, the goal in the following is to capture the most essential features of the expected detector response: threshold effects in detection, the finite accuracy and skew of reconstruction resolution functions, as well as limited flavor and charge identification capabilities. This does not invalidate the conclusions drawn from comparing the various analysis approaches.
\subsection{Detection Efficiency}
\begin{figure}[t]
    \centering
    \includegraphics[width=0.9\textwidth]{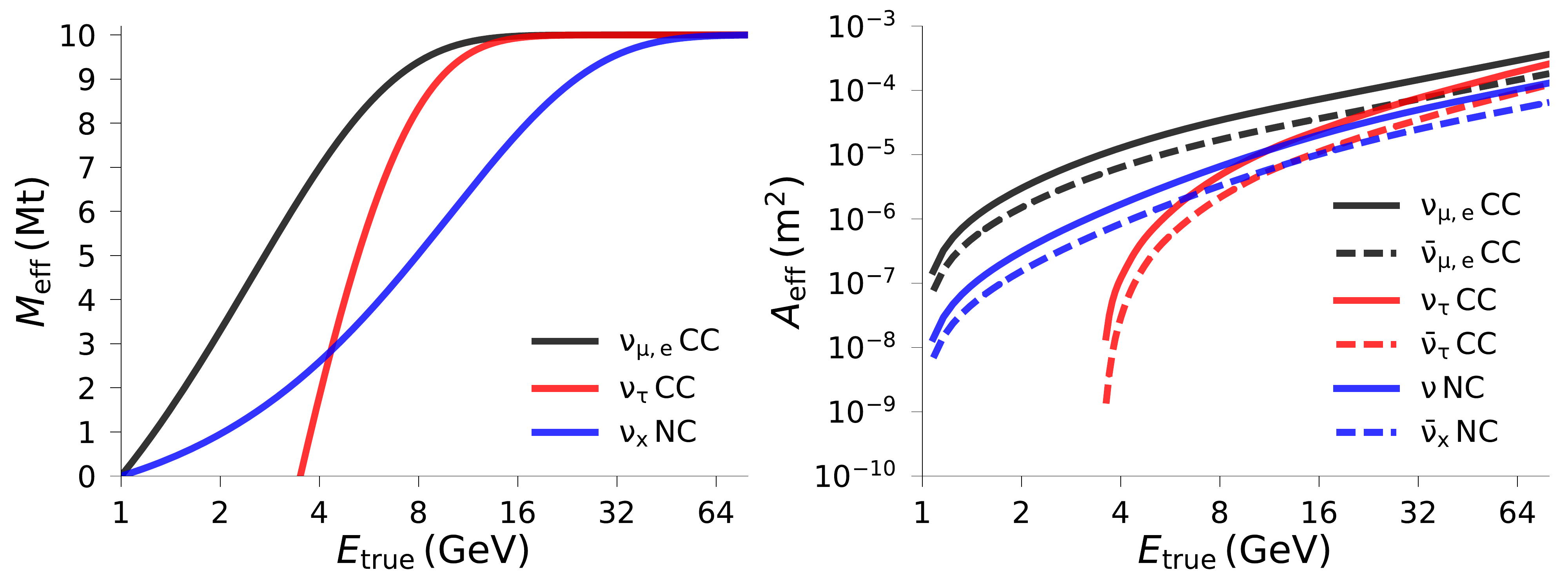}
    \caption{Effective masses (left) and areas (right) as a function of true neutrino energy for a generic toy detector with fiducial mass of $\SI{10}{\mega\tonne}$. The dependency of the effective masses on energy is given in Equation~(\ref{eq: MeffGen}). Cross sections are from Equation~(\ref{eq: xsecGen}), except for {\nutau} and {\nutaubar} interactions, which are interpolated from~\cite{Gazizov:2016dhn}. Effective masses are the same for neutrinos and anti-neutrinos. See text for details.}
    \label{fig: MeffAeff}
\end{figure}

We assume a detector of fiducial mass $M_\mathrm{fid} = \SI{10}{\rm megaton}$, with a neutrino detection energy threshold of $E_{\mathrm{th}}=\SI{1}{\giga\electronvolt}$ for all neutrino flavors and interaction channels apart from $\nu_\tau$ charged current (CC) interactions, where the intrinsic interaction threshold is higher, at $E_{\mathrm{th}}=\SI{3.5}{\giga\electronvolt}$. The detector's effective mass $M^\alpha_\mathrm{eff} = \rho_\mathrm{ice}V^\alpha_\mathrm{eff}$ for a given combination, $\alpha$, of flavor and interaction type, where $\rho_\mathrm{ice}$ is the ice density and $V^\alpha_\mathrm{eff}$ the detector's corresponding effective volume, exhibits a phenomenological dependence on true neutrino energy, {\Etrue}, asymptotically approaching $M_\mathrm{fid}$ according to an exponential function:
\begin{equation}
        M_\mathrm{eff}^{\alpha}(\Etrue) = M_\mathrm{fid}\times\left(1-e^{-k_{\alpha}\times(\Etrue/\si{\giga\electronvolt}-E_{\mathrm{th}}/\si{\giga\electronvolt})}\right) \text{for } \Etrue>E_{\mathrm{th}}\text{ .}\label{eq: MeffGen}
\end{equation}
We include three effective masses to cover all the neutrino interaction channels: one for {\nue}, {\nuebar}, {\numu}, and {\numubar} CC, one for {\nutau} and {\nutaubar} CC, and one for all NC channels. For the CC channels we choose $k_\alpha=0.4$, while for the NC channels the function rises more slowly, with $k_\alpha=0.1$.
The left panel of Figure~\ref{fig: MeffAeff} shows these dependencies for neutrino energies up to $\Etrue=\SI{80}{\giga\electronvolt}$. The detector can be roughly considered fully efficient ($M_\mathrm{eff} = M_\mathrm{fid}$) for all detection channels above $\Etrue \approx \SI{50}{\giga\electronvolt}$.

The $\nu$-$\bar{\nu}$ asymmetry---which is required to make the NMO measurement---will be introduced through differences in flux and cross sections, i.e., it will become apparent in the detector's effective area. The latter we obtain from the effective mass via the conversion

\begin{equation}
    A^\alpha_\mathrm{eff}(\Etrue) = \sigma_\alpha(\Etrue)\times n_\mathrm{ice}/\rho_\mathrm{ice}\times M^\alpha_\mathrm{eff}(\Etrue)\text{ ,}
    \label{eq: AeffGen}
\end{equation}
where $\sigma_\alpha$ is the total neutrino-nucleon cross section for a given flavor-interaction channel $\alpha$, $n_\mathrm{ice}\approx \SI{6e23}{\per\centi\metre\cubed}$ is the nucleon density in ice, and $\rho_\mathrm{ice} \approx \SI{0.92}{\gram\per\centi\metre\cubed}$ the mass density.

We also make some simplifying assumptions about the cross sections used in Equation~(\ref{eq: AeffGen}), in that we take {\nue} and {\numu} ({\nuebar} and {\numubar}) CC cross sections to be the same, as well as all $\nu_x$ ($\bar{\nu}_x$) NC cross sections. In addition, we model all the mentioned cross sections to rise strictly linearly with {\Etrue} above $\Etrue = \SI{1}{\giga\electronvolt}$~\cite{Formaggio:2013kya}:
\begin{equation}
    \sigma_\alpha(\Etrue)/\Etrue = c_\alpha\times \SI{e-38}{\centi\metre\squared\per\giga\electronvolt}\text{ ,}\label{eq: xsecGen}
\end{equation}
where we set
\begin{eqnarray}
    c_{\nu_{e,\mu\,{\rm CC}}} = 2 c_{\bar{\nu}_{e,\mu\, {\rm CC}}} = 0.70 \text{ ,}\\ 
    c_{\nu_{x\,{\rm NC}}} = 2 c_{\bar{\nu}_{x\, {\rm NC}}} = 0.25 \text{ .}
\end{eqnarray}
To obtain {\nutau} ({\nutaubar}) CC effective areas, we interpolate the corresponding neutrino-nucleon cross section curves given in~\cite{Gazizov:2016dhn}. All resulting effective areas as a function of neutrino energy are depicted in the right panel of Figure~\ref{fig: MeffAeff}. We take these to be invariant in azimuth, but universally introduce an arbitrary, smooth polynomial modification $M$ with the zenith angle dependency
\begin{equation}
M(x) = \frac{1}{20}(-x^3 + x^2 - x) + 1\qquad (x\equiv\CZtrue),
\end{equation}
which we normalize to unit area\footnote{ $A_\mathrm{eff}(\Etrue)$ is the average over the full sky, $\CZtrue \in [-1, +1]$.}.

\subsection{Reconstruction Resolutions}

\begin{figure}[ht]
  \centering
  \begin{minipage}[t]{0.48\textwidth}\vspace{0pt}
    \includegraphics[width=.95\textwidth]{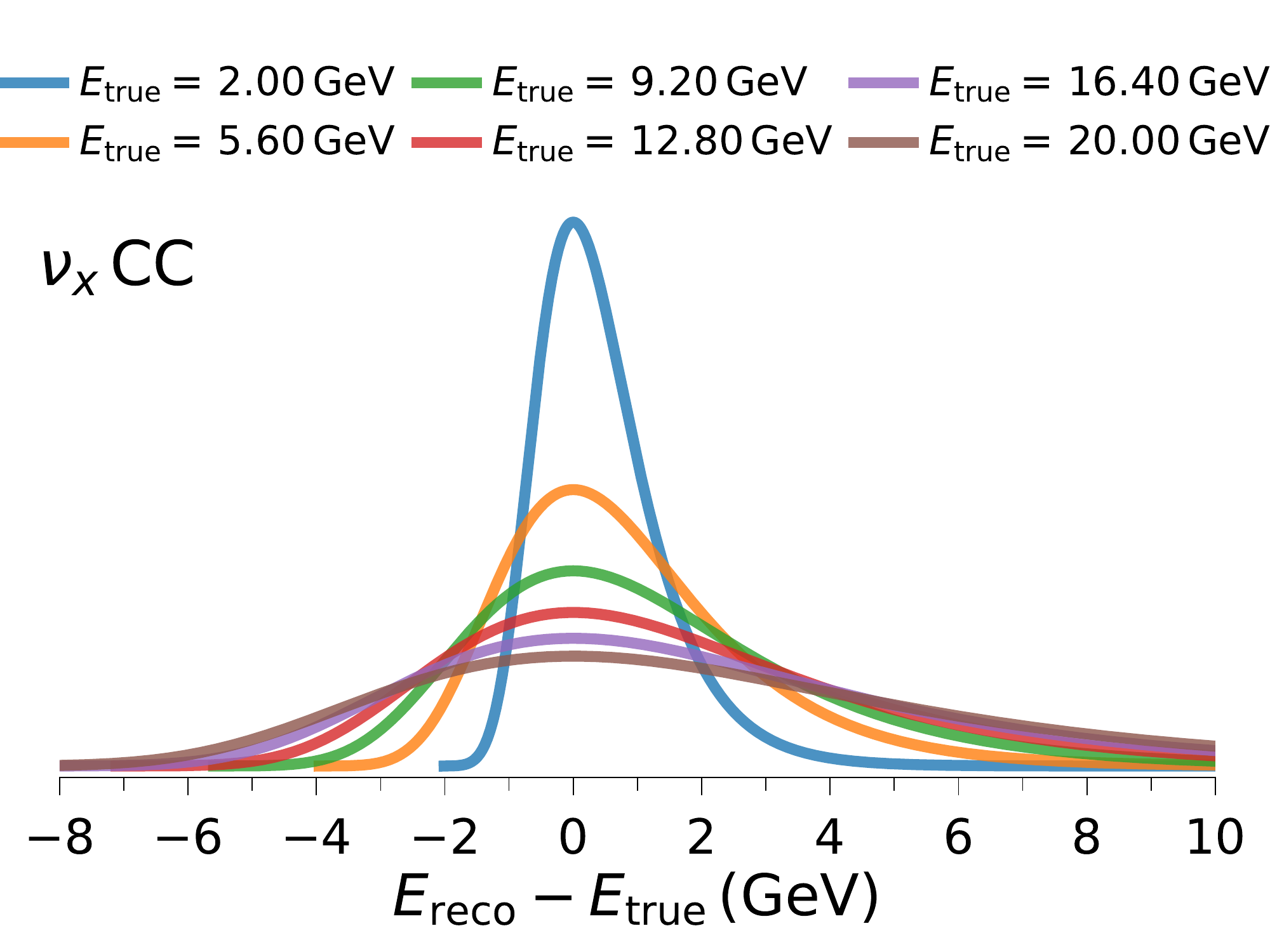}
    \caption{Example energy resolution functions (right-skewed Gumbel) used for all CC interactions, as given by Equation~(\ref{eq: reco cc}). The modes of the corresponding NC resolution functions are shifted by $-0.6E_\mathrm{true}$ with respect to the distributions depicted here.}
    \label{fig: Gumbelr}
  \end{minipage}
  \hfill
  \begin{minipage}[t]{0.48\textwidth}\vspace{0pt}
    \includegraphics[width=.95\textwidth]{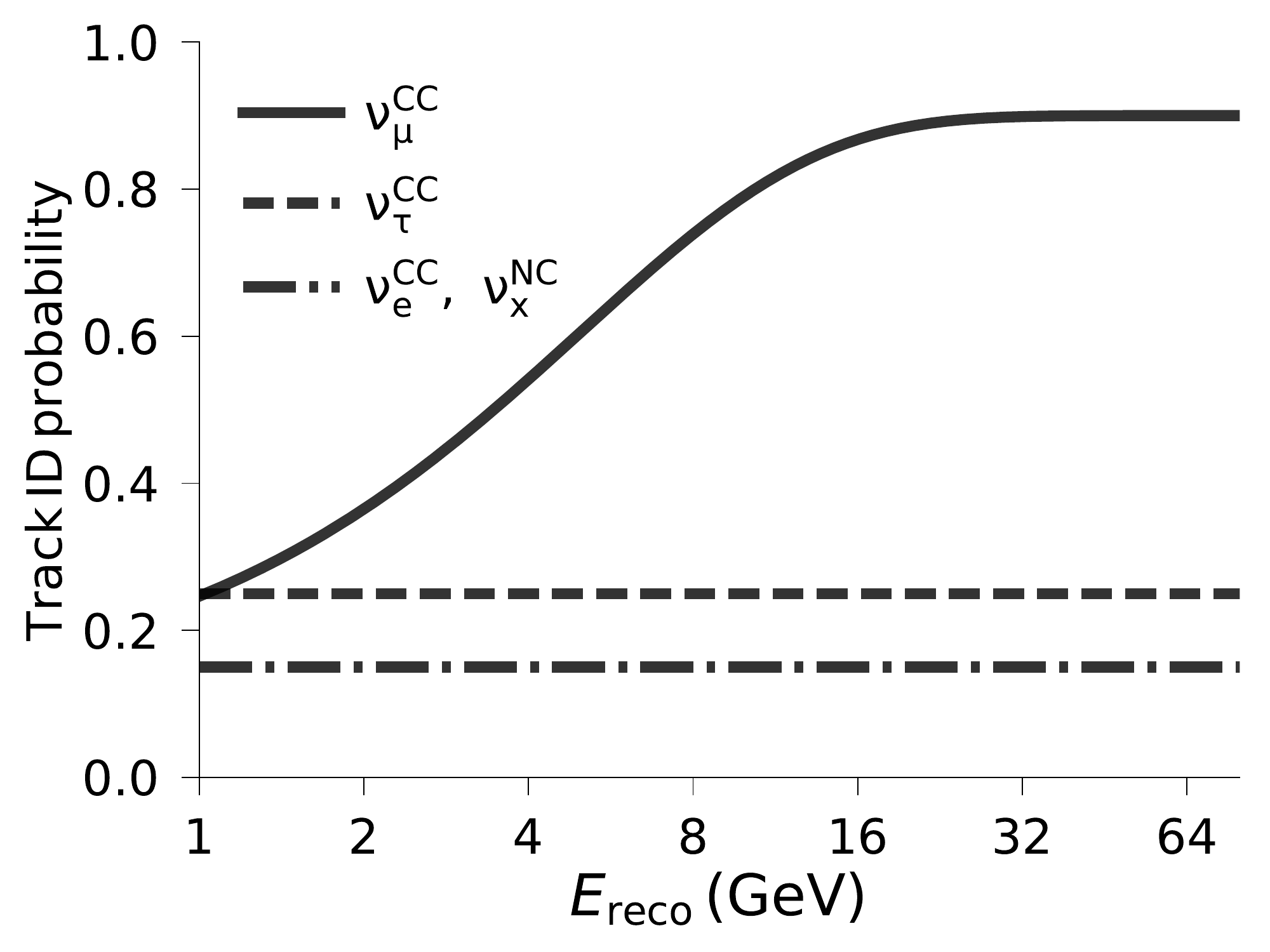}
    \caption{Event classification efficiencies implemented as functions of reconstructed neutrino energy. Shown is the probability to identify an event of a given type as ``track-like". Events are identified as ``cascade-like" with complementary probabilities.}
    \label{fig: PID}
  \end{minipage}

\end{figure}

\label{sec:resolutions}
Neutrino zenith resolutions with respect to $\cos \vartheta$ are represented by single Gaussian distributions. The distributions' parameters are taken as functions of {\Etrue} only. For each flavor and interaction channel, we assign a mean $\mu_{\Delta \cos\vartheta}(\Etrue) = 0$ across all energies, and a standard deviation of $\sigma_{\Delta\cos\vartheta}(\Etrue) = \frac{0.3}{\sqrt{\Etrue/\si{\giga\electronvolt}}}+0.05$.

Neutrino energy resolutions we describe using right-skewed Gumbel distributions. These are shifted and scaled by $\mu^\prime$ and $\sigma^\prime$ with respect to their standard form, via the transformation $x\rightarrow (x-\mu^\prime)/\sigma^\prime$. These parameters again only depend on {\Etrue}. The CC distributions are assumed identical for all flavors, and are shown in Figure~\ref{fig: Gumbelr}:
\begin{equation}
    \mu^{\prime\mathrm{CC}}_{\Delta E_\nu}(\Etrue) = 0, \quad
    \sigma^{\prime\mathrm{CC}}_{\Delta E_\nu}(\Etrue) = \left(\frac{0.4}{\sqrt{\Etrue/\si{\giga\electronvolt}}} + 0.1\right)\times \Etrue\text { .}
    \label{eq: reco cc}
\end{equation}
For NC interactions, we take a spread that scales with {\Etrue} in the same way $\sigma^{\prime\mathrm{CC}}_{\Delta E_\nu}$ does, but assume a non-zero mean due to the undetected energy carried away by the outgoing neutrino: $\mu^{\prime\mathrm{NC}}_{\Delta E_\nu}(\Etrue) = -0.6\Etrue$.

Note that each energy and cosine zenith residual distribution is renormalized such that its integral over the physical region ($\Delta E_\nu + \Etrue \geq 0$ and $-1 \leq (\Delta \cos\vartheta + {\CZtrue}) \leq 1$) yields $1$.
\subsection{Event Classification}
Correctly identifying few-$\si{\giga\electronvolt}$ CC muon neutrino interactions with relatively sparsely instrumented neutrino telescopes in water/ice is difficult mainly for two reasons. The track length of a near minimum ionizing muon is only on the order of a few meters, comparable to the extent of an electromagnetic cascade of the same energy. Also, photon scattering lengths similar to the horizontal spacing between photomultiplier tubes smear out the Cherenkov ring around the muon direction, which is otherwise observed at a specific angle with respect to the muon direction in the medium.

We take into account the muon neutrino CC (``track") identification efficiency $p^\mathrm{\mu,CC}_\mathrm{track}$ improving with (reconstructed) neutrino energy, $E_{\rm reco}$, by setting
\begin{equation}
    p^\mathrm{\mu,CC}_\mathrm{track} \equiv p^\mathrm{\mu,CC}_\mathrm{track}(E_\mathrm{reco}) = 0.9\times \left(1-e^{-0.2\times(E_\mathrm{reco}/\si{\giga\electronvolt}+0.6)}\right)\text{ .}
\end{equation}
This reflects the track length of the secondary muon increasing linearly with its energy, but also the possible production of a low-energy muon which cannot be distinguished from the accompanying hadronic cascade even for higher-energy muon neutrino CC interactions. All other (in)efficiencies are assumed to be constant:
\begin{eqnarray}
p^\mathrm{e,CC}_\mathrm{track}(E_\mathrm{reco}) = p^\mathrm{NC}_\mathrm{track}(E_\mathrm{reco}) = 0.15\text{ ,}\\
p^\mathrm{\tau,CC}_\mathrm{track}(E_\mathrm{reco}) = 0.25\text{ .}
\end{eqnarray}
These are shown in Figure~\ref{fig: PID}. The probability to identify any event as ``cascade-like" for a given reconstructed energy is just the complementary probability to that of the track identification.

When a toy MC event is subject to this classification, we assign it one of two discrete numbers---representative of either identification as track or cascade---with the above probabilities.

\section{Uncertainty in Significance}
\label{sec:signif}

Under the assumption that the test statistic $\mathcal{T}$ under two hypotheses $H_1$ and $H_2$ is normally distributed (with means $\mu_1$ and $\mu_2$ and with identical standard deviation $\sigma$), 
the number of standard deviations ($n_\sigma$) separating the two hypotheses can be written as $n_\sigma = |\mu_1 - \mu_2|/\sigma$  (corresponding to a one-sided hypothesis test and a one-sided conversion from p-value).
Sampling each distribution with $N_{\rm p}$ pseudo-experiments results in the following uncertainties for mean and standard deviation (see for example \cite{sivia1996data})

\begin{align}
\label{eq:uncert}
\Delta\mu = \frac{\sigma}{\sqrt{N_{\rm p}}}, \\
\Delta\sigma = \frac{\sigma}{\sqrt{2(N_{\rm p}-1)}}.
\end{align}
Since the combination of the quantities is linear, we can perform simple error propagation, so that the relative uncertainty in significance becomes (with $\oplus$ denoting sum in quadrature)
\begin{equation}
\label{eq:error}
\frac{\Delta n_\sigma}{n_\sigma} = \frac{\Delta\sigma}{\sigma} \oplus \frac{\Delta|\mu_1 - \mu_2|}{|\mu_1 - \mu_2|}.
\end{equation}
Using
\begin{equation}
\Delta|\mu_1 - \mu_2| = \Delta\mu_1 \oplus \Delta\mu_2 = \sqrt{\frac{2}{N_{\rm p}}}\sigma
\end{equation}
the second term simplifies to
\begin{equation}
\label{eq:simplified}
\frac{\Delta|\mu_1 - \mu_2|}{|\mu_1 - \mu_2|} = \sqrt{\frac{2}{N_{\rm p}}}\frac{\sigma}{|\mu_1 - \mu_2|} = \sqrt{\frac{2}{N_{\rm p}}}\frac{1}{n_\sigma}.
\end{equation}
Substituting Equations~(\ref{eq:simplified})~and~(\ref{eq:uncert}) into Equation~(\ref{eq:error}) yields
\begin{equation}
\frac{\Delta n_\sigma}{n_\sigma} = \frac{1}{\sqrt{2(N_{\rm p}-1)}} \oplus \sqrt{\frac{2}{N_{\rm p} n_\sigma^2}} = \sqrt{\frac{1}{2(N_{\rm p}-1)} + \frac{2}{N_{\rm p} n_\sigma^2}}.
\end{equation}
The absolute error on the number of standard deviations and its approximation for large $N_{\rm p}$ then follow immediately as
\begin{equation}
\Delta n_\sigma =  \sqrt{\frac{n_\sigma^2}{2(N_{\rm p}-1)} + \frac{2}{N_{\rm p}}} \xrightarrow{(N_{\rm p}>>1)} \frac{1}{\sqrt{N_{\rm p}}} \sqrt{\frac{n_\sigma^2}{2} + 2}.
\end{equation}


\bibliographystyle{elsarticle-num}

\bibliography{sample}

\begin{thebibliography}{10}
\expandafter\ifx\csname url\endcsname\relax
  \def\url#1{\texttt{#1}}\fi
\expandafter\ifx\csname urlprefix\endcsname\relax\def\urlprefix{URL }\fi
\expandafter\ifx\csname href\endcsname\relax
  \def\href#1#2{#2} \def\path#1{#1}\fi

\bibitem{upgrade_icrc:2019}
A.~Ishihara, {The IceCube Upgrade - Design and Science Goals}, in: {36th
  International Cosmic Ray Conference (ICRC 2019) Madison, Wisconsin, USA, July
  24-August 1, 2019}, 2019.
\newblock \href {http://arxiv.org/abs/1907.11699} {\path{arXiv:1907.11699}}.

\bibitem{TheIceCube-Gen2:2016cap}
M.~G. Aartsen, et~al., {PINGU: a vision for neutrino and particle physics at
  the South Pole}, J. Phys. G44~(5) (2017) 054006.
\newblock \href {http://arxiv.org/abs/1607.02671} {\path{arXiv:1607.02671}},
  \href {http://dx.doi.org/10.1088/1361-6471/44/5/054006}
  {\path{doi:10.1088/1361-6471/44/5/054006}}.

\bibitem{PINGU-LOI}
M.~G. Aartsen, et~al., {Letter of intent: the Precision IceCube Next Generation
  Upgrade (PINGU), (2017). }\href {http://arxiv.org/abs/1401.2046v2}
  {\path{arXiv:1401.2046v2}}.

\bibitem{Adrian-Martinez:2016fdl}
S.~Adrian-Martinez, et~al., {Letter of intent for KM3NeT 2.0}, J. Phys. G43~(8)
  (2016) 084001.
\newblock \href {http://arxiv.org/abs/1601.07459} {\path{arXiv:1601.07459}},
  \href {http://dx.doi.org/10.1088/0954-3899/43/8/084001}
  {\path{doi:10.1088/0954-3899/43/8/084001}}.

\bibitem{Glusenkamp:2019uir}
T.~Gl{\"u}senkamp, {A unified perspective on modified Poisson likelihoods for
  limited Monte Carlo data. }\href {http://arxiv.org/abs/1902.08831}
  {\path{arXiv:1902.08831}}.

\bibitem{Arguelles:2019izp}
C.~A. Arg{\"u}elles, A.~Schneider, T.~Yuan, {A binned likelihood for stochastic
  models}, JHEP 06 (2019) 030.
\newblock \href {http://arxiv.org/abs/1901.04645} {\path{arXiv:1901.04645}},
  \href {http://dx.doi.org/10.1007/JHEP06(2019)030}
  {\path{doi:10.1007/JHEP06(2019)030}}.

\bibitem{Barlow:2003cx}
R.~Barlow, {Introduction to statistical issues in particle physics},
  {Statistical problems in particle physics, astrophysics and cosmology.
  Proceedings, Conference, PHYSTAT 2003, Stanford, USA, September 8-11, 2003,}
  C030908 (2003) MOAT002.
\newblock \href {http://arxiv.org/abs/physics/0311105}
  {\path{arXiv:physics/0311105}}.

\bibitem{cowan}
G.~Cowan, Statistical Data Analysis, Oxford University Press, 1998.

\bibitem{l-bfgs-b:1995}
R.~H. Byrd, P.~Lu, J.~Nocedal, {A limited memory algorithm for bound
  constrained optimization}, SIAM J. Sci. Comput. 16 (1995) 1190--1208.

\bibitem{Cowan:2010js}
G.~Cowan, K.~Cranmer, E.~Gross, O.~Vitells, {Asymptotic formulae for
  likelihood-based tests of new physics}, Eur. Phys. J. C71 (2011) 1554,
  [Erratum: Eur. Phys. J.C73,2501(2013)].
\newblock \href {http://arxiv.org/abs/1007.1727} {\path{arXiv:1007.1727}},
  \href {http://dx.doi.org/10.1140/epjc/s10052-011-1554-0,
  10.1140/epjc/s10052-013-2501-z} {\path{doi:10.1140/epjc/s10052-011-1554-0,
  10.1140/epjc/s10052-013-2501-z}}.

\bibitem{Blennow:2014fk}
M.~Blennow, P.~Coloma, P.~Huber, T.~Schwetz, {Quantifying the sensitivity of
  oscillation experiments to the neutrino mass ordering}, JHEP 2014~(3) (2014)
  28.
\newblock \href {http://dx.doi.org/10.1007/JHEP03(2014)028}
  {\path{doi:10.1007/JHEP03(2014)028}}.

\bibitem{Cranmer:2000du}
K.~S. Cranmer, {Kernel estimation in high-energy physics}, Comput. Phys.
  Commun. 136 (2001) 198--207.
\newblock \href {http://arxiv.org/abs/hep-ex/0011057}
  {\path{arXiv:hep-ex/0011057}}, \href
  {http://dx.doi.org/10.1016/S0010-4655(00)00243-5}
  {\path{doi:10.1016/S0010-4655(00)00243-5}}.

\bibitem{scott}
D.~W. Scott, On optimal and data-based histograms, Biometrika 66~(3) (1979)
  605.
\newblock \href {http://dx.doi.org/10.1093/biomet/66.3.605}
  {\path{doi:10.1093/biomet/66.3.605}}.

\bibitem{Fukuda:1998mi}
Y.~Fukuda, et~al., {Evidence for oscillation of atmospheric neutrinos}, Phys.
  Rev. Lett. 81 (1998) 1562--1567.
\newblock \href {http://arxiv.org/abs/hep-ex/9807003}
  {\path{arXiv:hep-ex/9807003}}, \href
  {http://dx.doi.org/10.1103/PhysRevLett.81.1562}
  {\path{doi:10.1103/PhysRevLett.81.1562}}.

\bibitem{Ahmad:2001an}
Q.~R. Ahmad, et~al., {Measurement of the rate of $\nu_e+d \to p+p+e^-$
  interactions produced by $^8B$ solar neutrinos at the Sudbury Neutrino
  Observatory}, Phys. Rev. Lett. 87 (2001) 071301.
\newblock \href {http://arxiv.org/abs/nucl-ex/0106015}
  {\path{arXiv:nucl-ex/0106015}}, \href
  {http://dx.doi.org/10.1103/PhysRevLett.87.071301}
  {\path{doi:10.1103/PhysRevLett.87.071301}}.

\bibitem{Aharmim:2011vm}
B.~Aharmim, et~al., {Combined analysis of all three phases of solar neutrino
  data from the Sudbury Neutrino Observatory}, Phys. Rev. C88 (2013) 025501.
\newblock \href {http://arxiv.org/abs/1109.0763} {\path{arXiv:1109.0763}},
  \href {http://dx.doi.org/10.1103/PhysRevC.88.025501}
  {\path{doi:10.1103/PhysRevC.88.025501}}.

\bibitem{Capozzi:2018ubv}
F.~Capozzi, E.~Lisi, A.~Marrone, A.~Palazzo, {Current unknowns in the three
  neutrino framework}, Prog. Part. Nucl. Phys. 102 (2018) 48--72.
\newblock \href {http://arxiv.org/abs/1804.09678} {\path{arXiv:1804.09678}},
  \href {http://dx.doi.org/10.1016/j.ppnp.2018.05.005}
  {\path{doi:10.1016/j.ppnp.2018.05.005}}.

\bibitem{deSalas:2018bym}
P.~F. De~Salas, S.~Gariazzo, O.~Mena, C.~A. Ternes, M.~T\'{o}rtola, {Neutrino
  Mass Ordering from Oscillations and Beyond: 2018 Status and Future
  Prospects}, Front. Astron. Space Sci. 5 (2018) 36.
\newblock \href {http://arxiv.org/abs/1806.11051} {\path{arXiv:1806.11051}},
  \href {http://dx.doi.org/10.3389/fspas.2018.00036}
  {\path{doi:10.3389/fspas.2018.00036}}.

\bibitem{nufit4.0}
{NuFIT 4.0 (2018), www.nu-fit.org}.

\bibitem{Esteban:2018azc}
I.~Esteban, M.~C. Gonzalez-Garcia, A.~Hernandez-Cabezudo, M.~Maltoni,
  T.~Schwetz, {Global analysis of three-flavour neutrino oscillations:
  synergies and tensions in the determination of $\theta_{23}, \delta_{CP}$,
  and the mass ordering}, JHEP 01 (2019) 106.
\newblock \href {http://arxiv.org/abs/1811.05487} {\path{arXiv:1811.05487}},
  \href {http://dx.doi.org/10.1007/JHEP01(2019)106}
  {\path{doi:10.1007/JHEP01(2019)106}}.

\bibitem{Patrignani:2016xqp}
C.~Patrignani, et~al., {Review of particle physics}, Chin. Phys. C40~(10)
  (2016) 100001.
\newblock \href {http://dx.doi.org/10.1088/1674-1137/40/10/100001}
  {\path{doi:10.1088/1674-1137/40/10/100001}}.

\bibitem{Wolfenstein:1978ue}
L.~Wolfenstein, {Neutrino oscillations in matter}, Phys. Rev. D17 (1978) 2369.

\bibitem{Mikheev:1986wj}
S.~P. Mikheev, A.~Y. Smirnov, {Resonant amplification of neutrino oscillations
  in matter and solar neutrino spectroscopy}, Nuovo Cim. C9 (1986) 17--26.

\bibitem{Petcov:1986qg}
S.~T. Petcov, S.~Toshev, {Three neutrino oscillations in matter: analytical
  results in the adiabatic approximation}, Phys. Lett. B187 (1987) 120--126.
\newblock \href {http://dx.doi.org/10.1016/0370-2693(87)90083-9}
  {\path{doi:10.1016/0370-2693(87)90083-9}}.

\bibitem{Akhmedov:2005yj}
E.~K. Akhmedov, M.~Maltoni, A.~{\relax Yu}. Smirnov, {Oscillations of high
  energy neutrinos in matter: precise formalism and parametric resonance},
  Phys. Rev. Lett. 95 (2005) 211801.
\newblock \href {http://arxiv.org/abs/hep-ph/0506064}
  {\path{arXiv:hep-ph/0506064}}, \href
  {http://dx.doi.org/10.1103/PhysRevLett.95.211801}
  {\path{doi:10.1103/PhysRevLett.95.211801}}.

\bibitem{Abe:2011ts}
K.~Abe, et~al., {Letter of intent: the Hyper-Kamiokande experiment --- detector
  design and physics potential ---. }\href {http://arxiv.org/abs/1109.3262}
  {\path{arXiv:1109.3262}}.

\bibitem{Neyman-Pearson}
J.~Neyman, E.~S. Pearson, On the problem of the most efficient tests of
  statistical hypotheses, Philosophical Transactions of the Royal Society of
  London A: Mathematical, Physical and Engineering Sciences 231~(694-706)
  (1933) 289--337.
\newblock \href {http://dx.doi.org/10.1098/rsta.1933.0009}
  {\path{doi:10.1098/rsta.1933.0009}}.

\bibitem{Gonzalez-Garcia:2014bfa}
M.~C. Gonzalez-Garcia, M.~Maltoni, T.~Schwetz, {Updated fit to three neutrino
  mixing: status of leptonic CP violation}, JHEP 2014~(11) (2014) 52.
\newblock \href {http://dx.doi.org/10.1007/JHEP11(2014)052}
  {\path{doi:10.1007/JHEP11(2014)052}}.

\bibitem{nufit2.0}
{NuFIT 2.0 (2014), www.nu-fit.org}.

\bibitem{Honda:2015fha}
M.~Honda, M.~Sajjad~Athar, T.~Kajita, K.~Kasahara, S.~Midorikawa, {Atmospheric
  neutrino flux calculation using the NRLMSISE-00 atmospheric model}, Phys.
  Rev. D92~(2) (2015) 023004.
\newblock \href {http://arxiv.org/abs/1502.03916} {\path{arXiv:1502.03916}},
  \href {http://dx.doi.org/10.1103/PhysRevD.92.023004}
  {\path{doi:10.1103/PhysRevD.92.023004}}.

\bibitem{Barr:2006it}
G.~D. Barr, T.~K. Gaisser, S.~Robbins, T.~Stanev, {Uncertainties in atmospheric
  neutrino fluxes}, Phys. Rev. D74 (2006) 094009.
\newblock \href {http://arxiv.org/abs/astro-ph/0611266}
  {\path{arXiv:astro-ph/0611266}}, \href
  {http://dx.doi.org/10.1103/PhysRevD.74.094009}
  {\path{doi:10.1103/PhysRevD.74.094009}}.

\bibitem{Evans:2016obt}
J.~Evans, D.~G. Gamez, S.~D. Porzio, S.~S{\"o}ldner-Rembold, S.~Wren,
  {Uncertainties in atmospheric muon-neutrino fluxes arising from cosmic-ray
  primaries}, Phys. Rev. D95~(2) (2017) 023012.
\newblock \href {http://arxiv.org/abs/1612.03219} {\path{arXiv:1612.03219}},
  \href {http://dx.doi.org/10.1103/PhysRevD.95.023012}
  {\path{doi:10.1103/PhysRevD.95.023012}}.

\bibitem{PhysRevD.22.2718}
V.~Barger, K.~Whisnant, S.~Pakvasa, R.~J.~N. Phillips, Matter effects on
  three-neutrino oscillations, Phys. Rev. D22 (1980) 2718--2726.
\newblock \href {http://dx.doi.org/10.1103/PhysRevD.22.2718}
  {\path{doi:10.1103/PhysRevD.22.2718}}.

\bibitem{prob3}
R.~Wendell, {Prob3++} software for computing three flavor neutrino oscillation
  probabilities, \url{http://www.phy.duke.edu/~raw22/public/Prob3++/}, 2012.

\bibitem{Dziewonski1981297}
A.~M. Dziewonski, D.~L. Anderson, {Preliminary reference Earth model}, {Physics
  of the Earth and planetary interiors} 25~(4) (1981) 297 -- 356.
\newblock \href
  {http://dx.doi.org/http://dx.doi.org/10.1016/0031-9201(81)90046-7}
  {\path{doi:http://dx.doi.org/10.1016/0031-9201(81)90046-7}}.

\bibitem{Bulmahn:2010pg}
A.~Bulmahn, M.~H. Reno, {Secondary atmospheric tau neutrino production}, Phys.
  Rev. D82 (2010) 057302.
\newblock \href {http://arxiv.org/abs/1007.4989} {\path{arXiv:1007.4989}},
  \href {http://dx.doi.org/10.1103/PhysRevD.82.057302}
  {\path{doi:10.1103/PhysRevD.82.057302}}.

\bibitem{Wren:2018}
S.~Wren,
  \href{https://www.research.manchester.ac.uk/portal/en/theses/neutrino-mass-ordering-studies-with-icecubedeepcore(70414fde-3bef-4028-877b-5e1e86b2165d).html}{{Neutrino
  Mass Ordering Studies with IceCube-DeepCore}}, Ph.D. thesis, Manchester U.
  (2018).
\newline\urlprefix\url{https://www.research.manchester.ac.uk/portal/en/theses/neutrino-mass-ordering-studies-with-icecubedeepcore(70414fde-3bef-4028-877b-5e1e86b2165d).html}

\bibitem{Calland:2013vaa}
R.~G. Calland, A.~C. Kaboth, D.~Payne, {Accelerated event-by-event neutrino
  oscillation reweighting with matter effects on a GPU}, JINST 9 (2014) P04016.
\newblock \href {http://arxiv.org/abs/1311.7579} {\path{arXiv:1311.7579}},
  \href {http://dx.doi.org/10.1088/1748-0221/9/04/P04016}
  {\path{doi:10.1088/1748-0221/9/04/P04016}}.

\bibitem{Nickolls2008ScalablePP}
J.~Nickolls, I.~Buck, M.~Garland, K.~Skadron, Scalable parallel programming
  with cuda, 2008 IEEE Hot Chips 20 Symposium (HCS) (2008) 1--2.

\bibitem{Wallraff:2014qka}
M.~Wallraff, C.~Wiebusch, {Calculation of oscillation probabilities of
  atmospheric neutrinos using nuCraft}, Comput. Phys. Commun. 197 (2015)
  185--189.
\newblock \href {http://arxiv.org/abs/1409.1387} {\path{arXiv:1409.1387}},
  \href {http://dx.doi.org/10.1016/j.cpc.2015.07.010}
  {\path{doi:10.1016/j.cpc.2015.07.010}}.

\bibitem{PhysRevD.57.1977}
T.~K. Gaisser, T.~Stanev, Path length distributions of atmospheric neutrinos,
  Phys. Rev. D57 (1998) 1977--1982.
\newblock \href {http://dx.doi.org/10.1103/PhysRevD.57.1977}
  {\path{doi:10.1103/PhysRevD.57.1977}}.

\bibitem{abramson1982}
I.~S. Abramson, {On bandwidth variation in kernel estimates---a square root
  law}, Ann. Statist. 10~(4) (1982) 1217--1223.
\newblock \href {http://dx.doi.org/10.1214/aos/1176345986}
  {\path{doi:10.1214/aos/1176345986}}.

\bibitem{botev2010}
Z.~I. Botev, J.~F. Grotowski, D.~P. Kroese, {Kernel density estimation via
  diffusion}, Ann. Statist. 38~(5) (2010) 2916--2957.
\newblock \href {http://dx.doi.org/10.1214/10-AOS799}
  {\path{doi:10.1214/10-AOS799}}.

\bibitem{10.2307/2345597}
S.~J. Sheather, M.~C. Jones, A reliable data-based bandwidth selection method
  for kernel density estimation, Journal of the Royal Statistical Society.
  Series B (Methodological) 53~(3) (1991) 683--690.

\bibitem{Formaggio:2013kya}
J.~A. Formaggio, G.~P. Zeller, {From eV to EeV: neutrino cross sections across
  energy scales}, Rev. Mod. Phys. 84 (2012) 1307--1341.
\newblock \href {http://arxiv.org/abs/1305.7513} {\path{arXiv:1305.7513}},
  \href {http://dx.doi.org/10.1103/RevModPhys.84.1307}
  {\path{doi:10.1103/RevModPhys.84.1307}}.

\bibitem{Gazizov:2016dhn}
A.~Gazizov, M.~Kowalski, K.~S. Kuzmin, V.~A. Naumov, C.~Spiering,
  {Neutrino-nucleon cross sections at energies of megaton-scale detectors}, EPJ
  Web Conf. 116 (2016) 08003.
\newblock \href {http://arxiv.org/abs/1604.02092} {\path{arXiv:1604.02092}},
  \href {http://dx.doi.org/10.1051/epjconf/201611608003}
  {\path{doi:10.1051/epjconf/201611608003}}.

\bibitem{sivia1996data}
D.~Sivia, Data Analysis: A Bayesian Tutorial, Oxford University Press, 2006.

\end{thebibliography}

\end{document}